\documentclass[a4paper,12pt]{article}
\usepackage{amssymb}
\usepackage{tipa}
\usepackage{mathrsfs}
\usepackage{bbm}
\usepackage{epsf,amsmath}
\usepackage[dvips,usenames]{color}
\usepackage{graphicx,subfigure}
\usepackage{color}
\usepackage{colortbl}

\newlength{\dinwidth}
\newlength{\dinmargin}
\def\be{\begin{equation}}
\def\ee{\end{equation}}
\def\ba{\begin{eqnarray}}
\def\ea{\end{eqnarray}}

\begin{document}
\title{\bf Revisiting the Annihilation Corrections in Non-leptonic $\bar{B}_s^0$ Decays within QCD Factorization}
\author{Qin Chang$^{a,b}\footnote{changqin@htu.cn}$, Xiao-Wei Cui$^{a}$, Lin Han$^{b}$, Ya-Dong Yang$^{b,c}$\footnote{yangyd@iopp.ccnu.edu.cn}\\
{ $^a$\small Department of Physics, Henan Normal University,
Xinxiang, Henan 453007, P.~R. China}\\
{$^{b}$\small Institute of Particle Physics, Huazhong Normal
University, Wuhan,
Hubei  430079, P. R. China}\\
{ $^c$\small Key Laboratory of Quark \& Lepton Physics, Ministry of
Education, P. R. China}}
\date{}
\maketitle
\bigskip\bigskip
\maketitle \vspace{-1.5cm}
\begin{abstract}
\setlength{\baselineskip}{0.8\baselineskip}
Motivated by the  recent measurements of non-leptonic $\bar{B}_s^0$ decays by CDF and LHCb collaborations, especially the large ${\cal B}(\bar{B}_s^0\to\pi^+\pi^-)$, we revisit the hard spectator and annihilation corrections in $\bar{B}_s^0$ decays within QCD factorization approach with two schemes for the possible parameters for the known end-point divergence appeared in the estimation of the hard spectator and annihilation diagrams. The first one is to conservatively estimate the possible contributions by parameterization~(scheme I); another one uses an infrared finite gluon propagator~(scheme II) to  regulate the end-point singularity. In scheme~I, with the constraints from the measured $\bar{B}_s^0\to PP~(VV)$ decays, two~(four) restricted solutions of the parameters spaces are found. In scheme~II, we find that most of the theoretical predictions agree well with the experimental data with single parameter $m_{g}\sim 0.5{\rm GeV}$. However, within both schemes, ${\cal B}(\bar{B}_s^0\to\phi\phi)$ are always much larger than ${\cal B}(\bar{B}_s^0\to K^{\ast 0}\bar{K}^{\ast 0})$ in contrast to  the experimental results ${\cal B}(\bar{B}_s^0\to\phi\phi)\simeq {\cal B}(\bar{B}_s^0\to K^{\ast 0}\bar{K}^{\ast 0})$. It is noted that  the pattern ${\cal B}(\bar{B}_s^0\to\phi\phi) >{\cal B}(\bar{B}_s^0\to K^{\ast 0}\bar{K}^{\ast 0})$ also persists  in other theoretical framework,  thus  the present experimental result ${\cal B}(\bar{B}_s^0\to\phi\phi) \simeq {\cal B}(\bar{B}_s^0\to K^{\ast 0}\bar{K}^{\ast 0})$  rises a challenge to theoretical approaches for B non-leptonic decays. Further refined measurements and theoretical studies are  called for to resolve  such a possible anomaly.
\end{abstract}
\newpage
\section{Introduction}
The pure annihilation non-leptonic B meson decays, without the interference induced by spectator diagrams are very suitable for probing the strength of annihilation contribution and exploiting the related mechanism. Recently, CDF and LHCb Collaborations have reported the evidence of the pure annihilation decay $\bar{B}_s^0\to \pi^+\pi^-$, with a significance of $3.7 \sigma$ and $5.3\sigma$, respectively, 
\ba
{\cal B}(\bar{B}_s^0\to \pi^+\pi^-)&=&(0.57\pm0.15({\rm stat})\pm0.10 ({\rm syst}))\times10^{-6}\,,\quad {\rm CDF}~\cite{CDF}\label{CDFpipi}\\
{\cal B}(\bar{B}_s^0\to \pi^+\pi^-)&=&(0.98^{+0.23}_{-0.19}({\rm stat})\pm0.11 ({\rm syst}))\times10^{-6}\,,\quad {\rm LHCb}~\cite{LHCb}\label{LHCbpipi}
\ea
as well as the branching fraction of the pure annihilation decay $\bar{B}_d^0\to K^+K^-$,
\ba
{\cal B}(\bar{B}_d^0\to K^+K^-)&=&(0.23\pm0.10({\rm stat})\pm0.10 ({\rm syst}))\times10^{-6}\,,\quad {\rm CDF}~\cite{CDF}\label{CDFkk}\\
{\cal B}(\bar{B}_d^0\to K^+K^-)&=&(0.13^{+0.06}_{-0.05}({\rm stat})\pm0.07 ({\rm syst}))\times10^{-6}\,.\quad {\rm LHCb}~\cite{LHCb}\label{LHCbkk}
\ea
Averaging the experimental data Eqs.~(\ref{CDFpipi}) and (\ref{LHCbpipi}) ,  Heavy Flavor Averaging Group~(HFAG) gives
\be
{\cal B}(\bar{B}_s^0\to \pi^+\pi^-)=(0.73\pm0.14)\times10^{-6}\,.\quad {\rm HFAG}~\cite{HFAG}\label{HFAGpipi}
\ee
Averaging the experimental data Eqs.~(\ref{CDFkk}) and (\ref{LHCbkk}) roughly, we get
\be
{\cal B}(\bar{B}_d^0\to K^+K^-)=(0.16\pm0.08)\times10^{-6}\,.\label{Avrkk}
\ee

Theoretically, the pure annihilation non-leptonic B meson decays are expected much rare with a branching fractions at the level $10^{-7}$ or less due to the fact that the annihilation corrections are formally $\Lambda_{QCD}/m_b$ power suppressed. While, together with the chirally enhanced power corrections, they offer interesting probes  for the dynamical mechanism governing these decays and exploration of CP violations, and therefore attract much more attention recently~\cite{ZhuGH,relaRef,LvCD,XiaoZJ}. Unfortunately,  in collinear factorization approach, the calculation of annihilation corrections always suffers from end-point divergence. Within the perturbative QCD~(pQCD) approach~\cite{KLS}, such divergence is regulated by introducing the parton transverse momentum $k_{T}$ at expense of modeling  additional $k_{T}$ dependence of meson distribution functions, and present a large complex annihilation corrections~\cite{LvCD,KLS}. The most recent renewed pQCD estimations of ${\cal B}(\bar{B}_s^0\to \pi^+\pi^-)$ and ${\cal B}(\bar{B}_d^0\to K^+K^-)$~\cite{XiaoZJ} are in good agreement with the CDF and LHCb measurements, however, a systematic examination combined with other correlated decays in the same framework is not available yet. In the soft-collinear effective theory~(SCET)~\cite{scet}, the annihilation diagrams are factorable and real~\cite{scetAnni} to the leading power of ${\cal O}(\alpha_{s}(m_{b})\Lambda_{QCD}/m_{b})$. 

In the QCD factorization approach (QCDF)~\cite{Beneke1}, there are mainly two ways to deal with the end-point singularity in weak annihilation calculation: (i) scheme~I, parameterization in a model independent way~\cite{Beneke2} with at least two phenomenological parameters introduced, for example $X_A=\int^{1}_{0}dy/y=\mathrm{ln}(m_b/\Lambda_h) (1+\rho_A e^{i\phi_{A}})$; (ii) scheme~II, using the infrared finite gluon propagator~\cite{YYgluon,Chang1}, for example $1/k^2\to1/(k^2-M_g(k^2)+i\epsilon)$. 

As a popular way, the scheme~I is widely used in the theoretical calculations~\cite{Beneke2,Beneke3,Cheng1,Cheng2}. Fitting to the data of $B_{u,d}\to PP$ decays,   a favored parameter value choice ``Scenario S4" is obtained in Ref.~\cite{Beneke2}: $\rho_A^{u,d}(PP)\sim1$ and $\phi_{A}^{u,d}(PP)\sim-55^{\circ}$, which leads to the prediction
\be
{\cal B}(\bar{B}_d^0\to K^+K^-)=0.070\times10^{-6}\,.
\ee
Assuming the default values of $\rho_A(PP)$ and $\phi_{A}(PP)$ in $B_s$ decays are similar to that in $B_{u,d}$ decays, Cheng {\it et al.} give the prediction ~\cite{Cheng2}
\be
{\cal B}(\bar{B}_s^0\to \pi^+\pi^-)=(0.26^{+0.00+0.10}_{-0.00-0.09})\times10^{-6}\,.
\ee
It is noted that above QCDF predictions are significantly smaller than the measurements Eqs.~(\ref{HFAGpipi}) and (\ref{Avrkk}). Especially, the default value ${\cal B}(\bar{B}_s^0\to \pi^+\pi^-)=0.26\times10^{-6}$  is about $3.4\sigma$ lower than the experimental data $(0.73\pm0.14)\times10^{-6}$, which implies possible much larger annihilation contributions in $B_s$ decays than previous prospect. Using the CDF results in Eq.~(\ref{CDFpipi}) solely,  a detail study about such topic has been performed by Zhu~\cite{ZhuGH}. Assuming universal values of $\rho_A(PP)$ and $\phi_{A}(PP)$ for $B_d$ and $B_s$ decays, it is found that QCDF is hardly to provide results in agreement with all of the well measured $B\to PP$ decays. Then, if the recent measurement of LHCb in Eq.~(\ref{LHCbpipi}) is considered, the tension would be further enlarged, which may  imply the parameters $\rho_A$ and $\phi_{A}$ are non-universal in $B_d$ and $B_s$ decays. So, it is worthy to fit their values with available data of $B_d$ and $B_s$ decays, respectively, and update the QCDF predictions. 

Within the scheme~II, the formula of annihilation corrections for $B\to PP$ and $PV$ decays have been given in Ref.~\cite{Chang1}. In this scheme, with the only one input parameter effective gluon mass scale $m_g=0.50\pm0.05{\rm GeV}$, the theoretical predictions of the observables for $B_{u,d}\to\pi K,\rho K$ and $\pi K^{\ast}$ decays are consistent with the experimental data~\cite{Chang1}.  So, it is deserved  to check if its predictions for the pure annihilation decays are in agreement with  the same effective gluon mass scale parameter. Furthermore, the pure annihilation $B_{d,s}\to VV$ decays, which involve more observables, may play an important role to test the methods of the end-point singularity regulation. So, in this paper, we calculate the annihilation corrections related to $B_{d,s}\to VV$ decays with the infrared finite gluon propagator.  

In Section~2, we briefly review the annihilation contributions within QCDF.  In Sections~3 and 4, with schemes~I and II for the end-point divergence regulation, we revisit $\bar{B}_s^0\to PP\,,PV$ and $VV$ decay modes, respectively. In our evaluations, the pure annihilation $B_s$ non-leptonic decays and the related well measured ones are examined  simultaneously. Section~5 contains our conclusions. Some amplitudes of $\bar{B}_s^0$ decays and the theoretical input parameters are summarized in Appendix~A and B, respectively.

\section{Brief review of the annihilation corrections within QCDF}
In the Standard Model~(SM), the effective weak Hamiltonian responsible for $b\to p$ transitions is given as~\cite{Buchalla:1996vs}
\begin{eqnarray}\label{eq:eff}
 {\cal H}_{\rm eff} &=& \frac{G_F}{\sqrt{2}} \biggl[V_{ub}
 V_{up}^* \left(C_1 O_1^u + C_2 O_2^u \right) + V_{cb} V_{cp}^* \left(C_1
 O_1^c + C_2 O_2^c \right) - V_{tb} V_{tp}^*\, \big(\sum_{i = 3}^{10}
 C_i O_i \big. \biggl. \nonumber\\
 && \biggl. \big. + C_{7\gamma} O_{7\gamma} + C_{8g} O_{8g}\big)\biggl] +
 {\rm h.c.},
\end{eqnarray}
where $V_{qb} V_{qs}^{\ast}$~($q=u, c$ and $t$) are products of the Cabibbo-Kobayashi-Maskawa~(CKM) matrix elements, $C_{i}$ the Wilson coefficients, and $O_i$ the relevant four-quark operators .

With the effective weak Hamiltonian Eq.~(\ref{eq:eff}), the QCDF  has been fully developed and extensively employed to calculate the hadronic B meson decays. The basic theoretical framework of $B_{u,d,s}\to PP,PV$ and $VV$ decays could be found in Refs.~\cite{Beneke1,Beneke2,Beneke3,Cheng1,Cheng2}. In this paper, we adopt the same convention and formula given in Refs.~\cite{Beneke2,Beneke3}, except for some corrections pointed out by Ref.~\cite{XQLi}. It is noted that the strength and associated strong-interaction phase of annihilation corrections and hard-spectator scattering contributions are numerically important to evaluate the branching ratios, the CP asymmetry and the polarization observables. Unfortunately, such power correction terms always suffer from the endpoint divergences, which violate the factorization. To probe their possible effects conservatively, the endpoint divergent integrals are treated as signs of infrared sensitive contribution and  usually parameterized by~\cite{Beneke2,Beneke3}, 
\begin{equation}\label{treat-for-anni}
\int_0^1 \frac{dx}{x}\, \to X_A\,, \qquad \int_0^1dx \frac{\textmd{ln}x}{x}\,\to -\frac{1}{2}(X_A)^2\,,\qquad \int_0^1 \frac{dx}{x^2}\,\to X_L\,,
 \end{equation}
where,
\be
X_A =(1+\rho e^{i\phi}) \ln\frac{m_B}{\Lambda_h}\,,\qquad X_L =(1+\rho e^{i\phi}) \frac{m_B}{\Lambda_h}
\ee
with $\Lambda_h$ being a typical scale of order $0.5{\rm GeV}$, and $\rho$, $\phi$ being unknown real parameters.  $X_H$ is treated in the same manner. The different choices of the parameters space of $\rho$ and $\phi$ correspond to various scenarios, which have been thoroughly discussed in Refs.~\cite{Beneke2,Beneke3,Cheng1,Cheng2}. 

Fitting the fruitful experimental measurements of $B_{u,d}\to PP, PV$ and $VP$  decays, a favored scenarios S4 is obtained in Ref.~\cite{Beneke2}. Furthermore, the fitted $\rho$ and $\phi$ for  $B_{u,d}\to VV$ decays are also given in Ref.~\cite{Beneke3,Cheng1}. Their results are summarized as 
\ba
\label{S4BdPP}
&&\rho_d^{PP}=1\,,\quad \phi_d^{PP}=-55^{\circ}\,;\\
\label{S4BdPV}
&&\rho_d^{PV}=1\,,\quad \phi_d^{PV}=-20^{\circ}\,,\quad\rho_d^{VP}=1\,,\quad \phi_d^{VP}=-70^{\circ}\,;\\
&&\rho_d^{\rho K^{\ast},K^{\ast}\bar{K}^{\ast}}=0.78\,,\quad \phi_d^{\rho K^{\ast},K^{\ast}\bar{K}^{\ast}}=-43^{\circ}\,;
\rho_d^{\phi K^{\ast},K^{\ast}\omega}=0.65\,,\quad \phi_d^{\phi K^{\ast},K^{\ast}\omega}=-53^{\circ}\,.
\ea
Assuming the default values of $\rho_A$ and $\phi_A$ in the $B_s$ decays are similar to that in $B_{u,d}$ decays,  Ref.~\cite{Cheng2} takes the values 
\ba
&&\rho_s^{PP}=1\,,\quad \phi_s^{PP}=-55^{\circ}\,;\label{S4PP}\\
&&\rho_s^{PV}=0.85\,,\quad \phi_s^{PV}=-30^{\circ}\,,\quad\rho_s^{VP}=0.9\,,\quad \phi_s^{VP}=-65^{\circ}\,;\label{S4PV}\\
&&\rho_s^{VV}=0.70\,,\quad \phi_s^{VV}=-55^{\circ}\label{S4VV}\,,
\ea
as the inputs for the $B_s$ decays. In this paper, we denote above parameter space as ``scenarios $\overline{S4}$" for convenience. It is noted that some non-leptonic $\bar{B}^0_s$ decays have been well measured in recent years, such as $\bar{B}_s^0\to\pi^+\pi^-$, $\pi^-K^+$, $K^-K^+$, $K^{\ast 0}\bar{K}^{\ast 0}$ and $\phi\phi$ decays. So, it is worth to check above parameter values and refit them with the updated data of $B_s$ decays. Furthermore, without the interference induced by  spectator diagrams,  the pure annihilation non-leptonic $B_s$ meson decays,  such as $\bar{B}^0_s\to\pi\pi$, $\rho\pi$ and $\rho\rho$ decays, are very suitable for probing the strength of the annihilation corrections and related mechanism. So, in this paper, we mainly pay our attention to such two types of  $B_s$ decays. 

\section{$\bar{B}_s^0\to PP$ and $PV$ decay modes}
\subsection{Within Scheme I}
With the annihilation parameters of scenarios $\overline{S4}$ for $\bar{B}_s^0\to PP$ and $PV$ decays given by Eqs.~(\ref{S4PP}) and (\ref{S4PV}), and the other input parameters listed in Appendix B, the predictions for the observables of pure annihilation decays $\bar{B}_s^0\to \pi\pi\,,\rho\pi$ and the well measured  decays $\bar{B}_s^0\to\pi^-K^+\,,K^-K^+$ are given in the third column of Table \ref{PPPV}. The theoretical uncertainties are mainly induced by the three parts: quark masses , CKM elements and decay constants, form factors. We first scan randomly the points in the allowed ranged of the input parameters of the three parts, respectively, and then add errors in quadrature.    

\begin{table}[t]
 \begin{center}
 \caption{The numerical results for the branching fractions $[\times10^{-6}]$ and the direct CP violations $[\times10^{-2}]$ of $\bar{B}_s^0\to\pi\pi\,,\rho\pi\,,\rho\rho\,,\pi^-K^+$ and $K^-K^+$ decays in each scenarios.}
 \label{PPPV}
 \vspace{0.5cm}
 \small
 \doublerulesep 0.1pt \tabcolsep 0.05in
 \begin{tabular}{lccccccccccc} \hline \hline
                                                                 &Exp                  &\multicolumn{3}{c}{Scheme I}                                                                                              &\multicolumn{1}{c}{Scheme II}\\
                                                                 &                        & $\overline{S4}$                                        & $\rm S^{PP}A$                       & $\rm S^{PP}B$                     &$m_g=0.48{\rm GeV}$      \\ \hline
 ${\cal B}(\bar{B}_s^0\to\pi^+\pi^-)$    &$0.73\pm0.14$&$0.21^{+0.05}_{-0.04}$     &$0.69^{+0.16}_{-0.16}$          &$0.66^{+0.17}_{-0.15}$      &$0.50^{+0.11}_{-0.10}$\\
 ${\cal B}(\bar{B}_s^0\to\pi^0\pi^0)$    &---                    &$0.10^{+0.02}_{-0.02}$     &$0.34^{+0.08}_{-0.08}$          &$0.33^{+0.08}_{-0.07}$      &$0.25^{+0.05}_{-0.05}$ \\
 ${\cal B}(\bar{B}_s^0\to\pi^+\rho^-)$  &---                    &$0.010^{+0.002}_{-0.002}$&$0.032^{+0.008}_{-0.007}$ &$0.036^{+0.009}_{-0.008}$   &$0.028^{+0.007}_{-0.006}$\\
 ${\cal B}(\bar{B}_s^0\to\pi^-\rho^+)$  &---                    &$0.011^{+0.003}_{-0.002}$&$0.046^{+0.013}_{-0.011}$ &$0.019^{+0.005}_{-0.004}$  &$0.028^{+0.007}_{-0.006}$\\
 ${\cal B}(\bar{B}_s^0\to\pi^0\rho^0)$  &---                   &$0.010^{+0.002}_{-0.002}$&$0.037^{+0.010}_{-0.008}$  &$0.025^{+0.006}_{-0.006}$  &$0.028^{+0.007}_{-0.006}$\\
  \hline
  $A_{CP}(\bar{B}_s^0\to\pi^+\pi^-)$      &---                  &$0$                                        &$0$                                         &$0$                                                                        &$0$\\
 $A_{CP}(\bar{B}_s^0\to\pi^0\pi^0)$     &---                  &$0$                                        &$0$                                           &$0$                                                                         &$0$ \\
 $A_{CP}(\bar{B}_s^0\to\pi^+\rho^-)$   &---                  &$-12^{+1}_{-1}$                  &$-30^{+3}_{-3}$                   &$-15^{+1}_{-1}$                                                 &$0$ \\
 $A_{CP}(\bar{B}_s^0\to\pi^-\rho^+)$   &---                 &$11^{+1}_{-1}$                    &$21^{+2}_{-2}$                     &$30^{+3}_{-3}$                                                     &$0$\\
 $A_{CP}(\bar{B}_s^0\to\pi^0\rho^0)$  &---                    &$0$                                       &$0$                                           &$0$                                                                 &$0$ \\
 \hline \hline
  ${\cal B}(\bar{B}_s^0\to\pi^-K^+)$     &$5.0\pm1.1$   &$5.9^{+0.9}_{-0.7}$           &$5.4^{+0.9}_{-0.7}$                &$5.3^{+1.0}_{-0.8}$               &$6.2^{+0.9}_{-0.7}$\\
  ${\cal B}(\bar{B}_s^0\to K^-K^+)$     &$25.4\pm3.7$ &$21.9^{+3.9}_{-3.8}$         &$23.8^{+6.1}_{-5.8}$              &$27.1^{+7.5}_{-6.6}$             &$27.8^{+5.2}_{-5.1}$\\
   \hline
 $A_{CP}(\bar{B}_s^0\to\pi^-K^+)$  &$39\pm17$  &$19^{+3}_{-3}$                         &$56^{+7}_{-8}$                      &$42^{+33}_{-19}$                   &$32^{+4}_{-5}$ \\
  $A_{CP}(\bar{B}_s^0\to K^-K^+)$    &---              &$-8^{+1}_{-1}$                        &$-22^{+2}_{-4}$                      &$-6^{+4}_{-33}$                      &$-11^{+2}_{-1}$\\
 \hline \hline
 \end{tabular}
 \end{center}
 \end{table}

Our QCDF results of scenarios $\overline{S4}$ listed in Tables~\ref{PPPV} are consistent with the former predictions given in Refs.~\cite{Beneke2,Cheng2}, and the small difference is mainly induced by the different input values and some corrections mentioned above. One may find most of the predictions agree well with the experimental measurements. However, we again find the theoretical prediction ${\cal B}(\bar{B}_s^0\to\pi^+\pi^-)\sim0.21\times10^{-6}$, which agrees well with the other theoretical results such as $\sim0.26\times10^{-6}$ in Ref.~\cite{Cheng2} and $\sim0.155\times10^{-6}$ in Ref.~\cite{Beneke2}, is about $3.7\sigma$ lower than the average of experimental data $(0.73\pm0.14)\times10^{-6}$. 

\begin{figure}[t]
\begin{center}
\subfigure[]{\includegraphics [width=7cm]{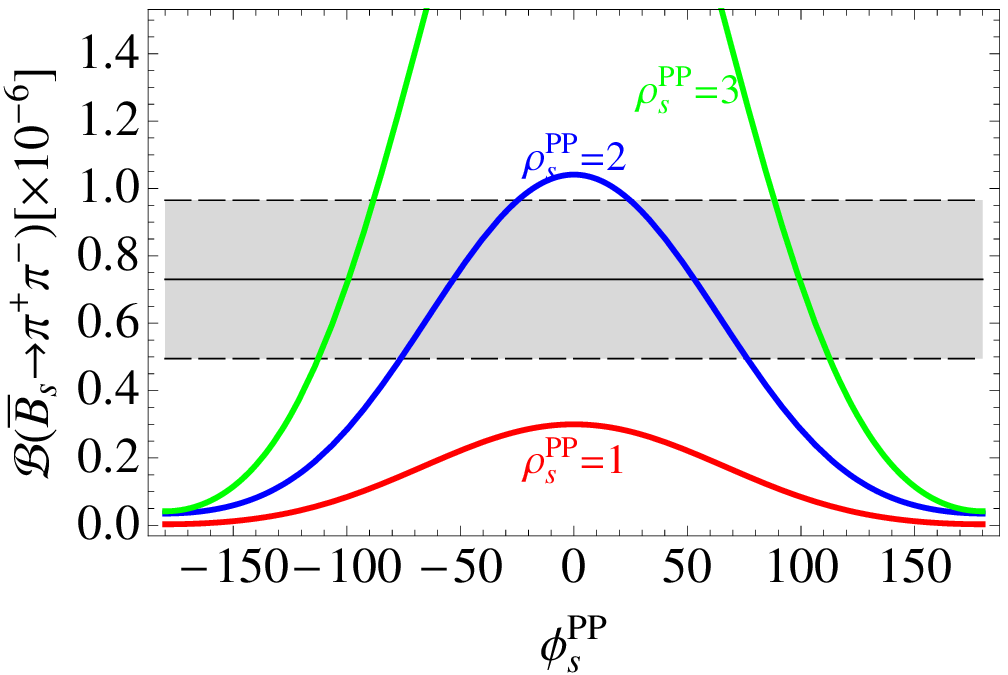}}
~~~~~~~~~~\subfigure[]{\includegraphics [width=7cm]{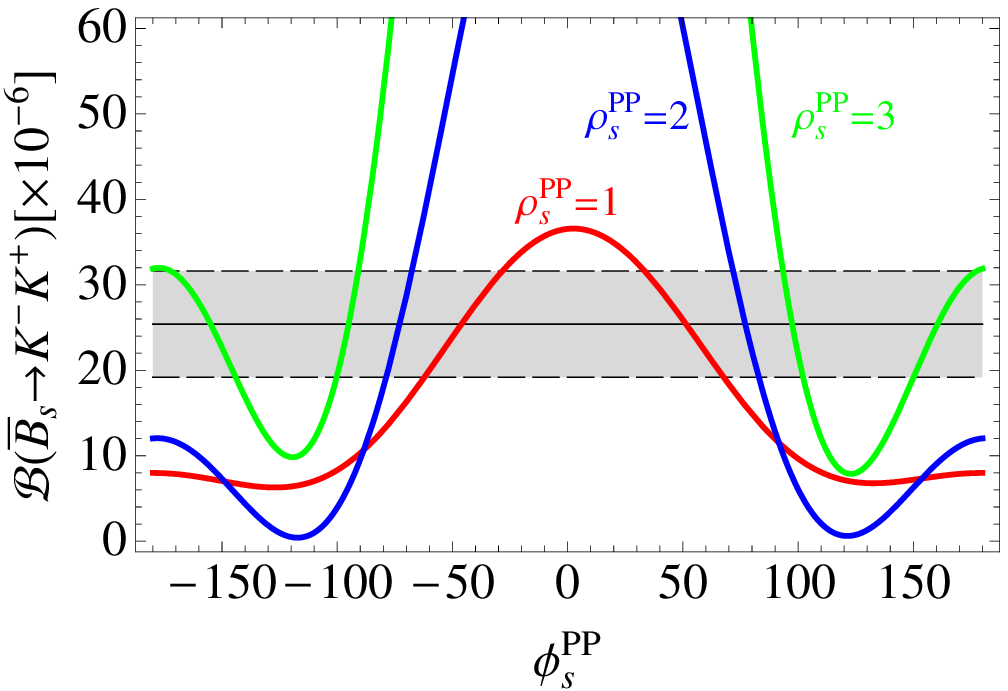}}\\
\subfigure[]{\includegraphics [width=7cm]{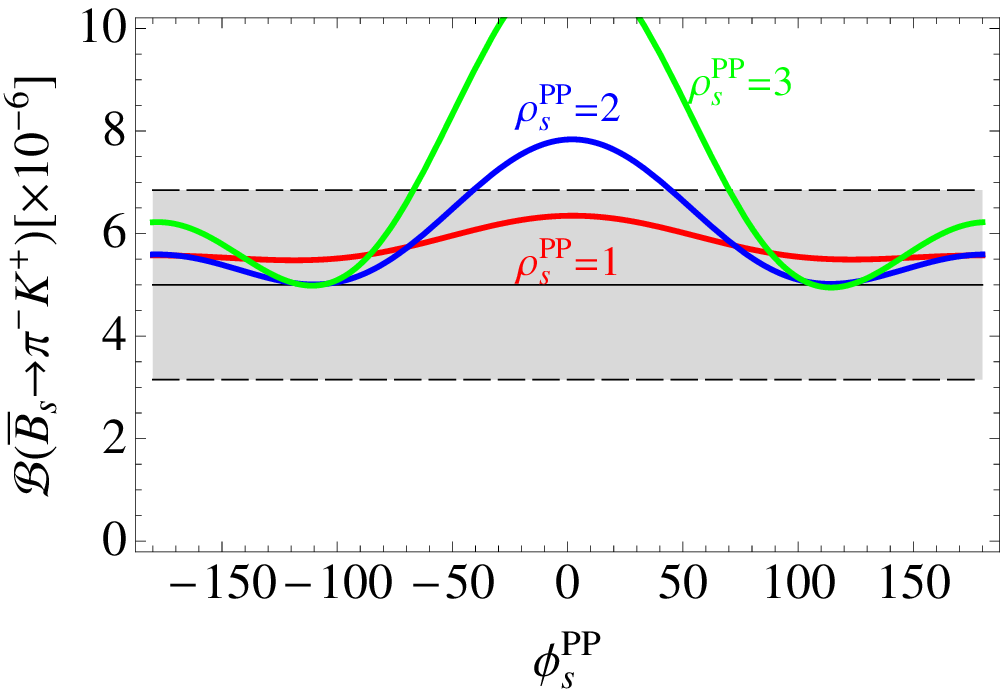}}
~~~~~~~~~~\subfigure[]{\includegraphics [width=7.1cm]{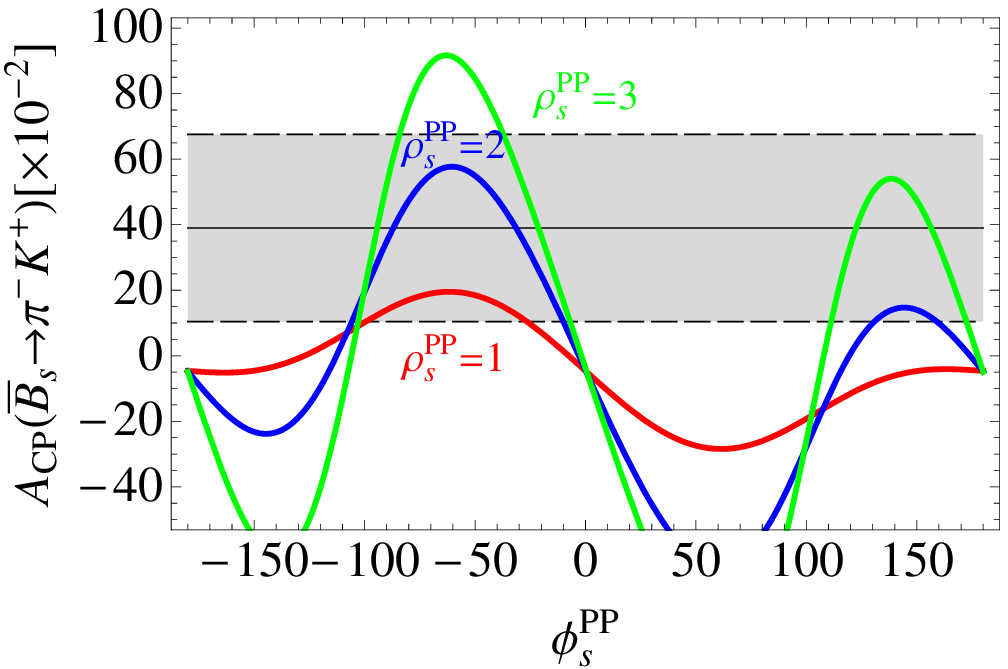}}\\
\centerline{\parbox{16cm}{\caption{\label{obsphiPP}\small The dependence of ${\cal B}(\bar{B}_s^0\to \pi^+\pi^-\,,K^+K^-\,,\pi^-K^+)$  and $A_{CP}(\bar{B}_s^0\to \pi^-K^+)$ on the phases $\phi_{s}^{PP}$ with different $\rho_{s}^{PP}$ values. The dashed lines correspond to the error bars~($1.68\sigma$).  }}}
\end{center}
\end{figure}

The Fig.~\ref{obsphiPP} shows the dependence of the measured observables of  $\bar{B}_s^0\to PP$ decays on the phase $\phi_{s}^{PP}$ with different $\rho_{s}^{PP}$ values. From Fig.~\ref{obsphiPP}~(a), one may easily find that the annihilation correction with the nominal annihilation parameter value $\rho_{s}^{PP}\sim1$ is hardly to account for the measured large ${\cal B}(\bar{B}_s^0\to\pi^+\pi^-)$ within errors, and a larger $\rho_{s}^{PP}$ is required. For the other measured observables, as Figs.~\ref{obsphiPP}~(b), (c) and (d) show, the large annihilation correction is not essential, and is allowed.  So, it is worthy to evaluate the exact values of the annihilation parameters with the constraints from the available experimental information of $\bar{B}_s^0\to PP$ decays.

\begin{figure}[ht]
\begin{center}
\subfigure[]{\includegraphics [width=5cm]{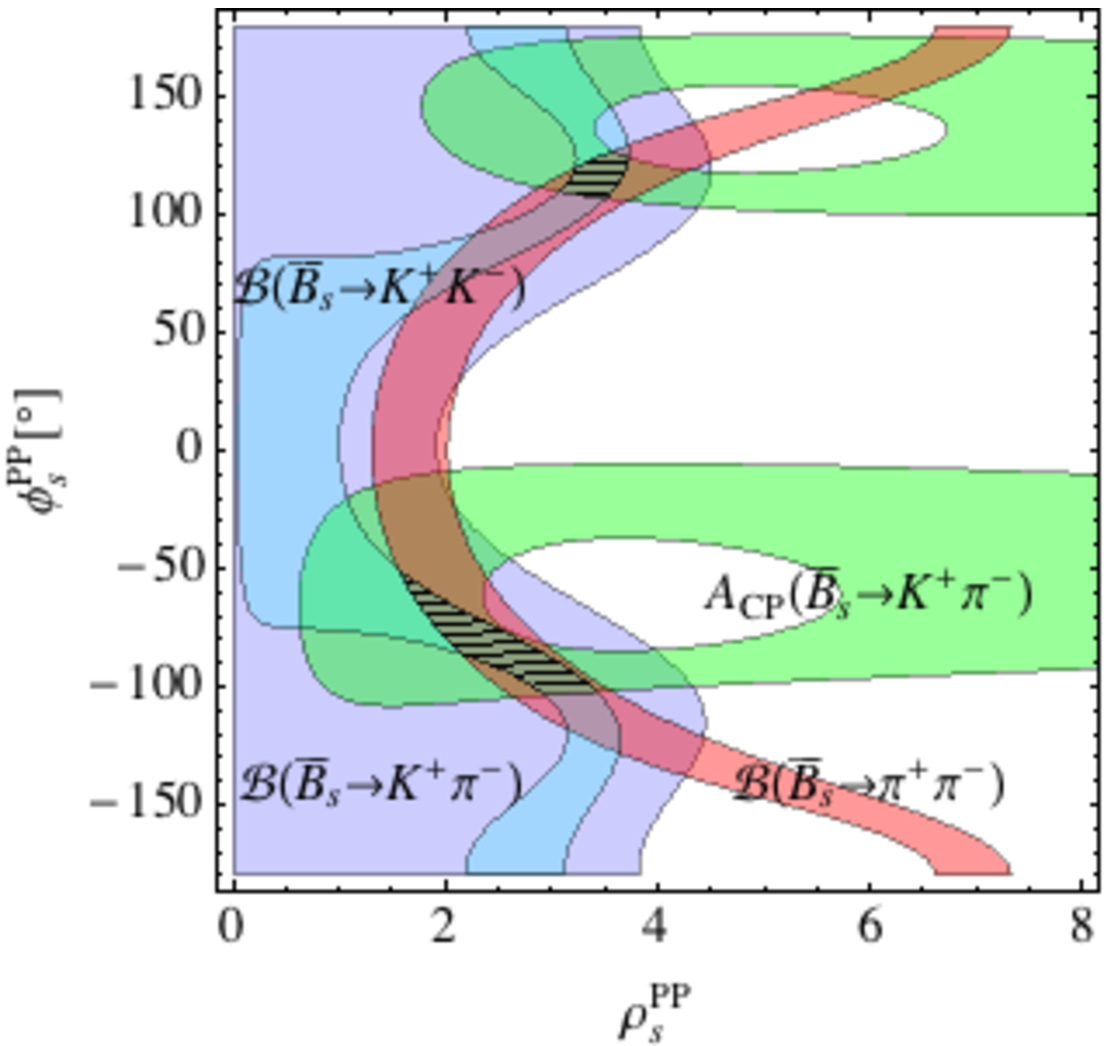}}
~~~~~~~~~~\subfigure[]{\includegraphics [width=5cm]{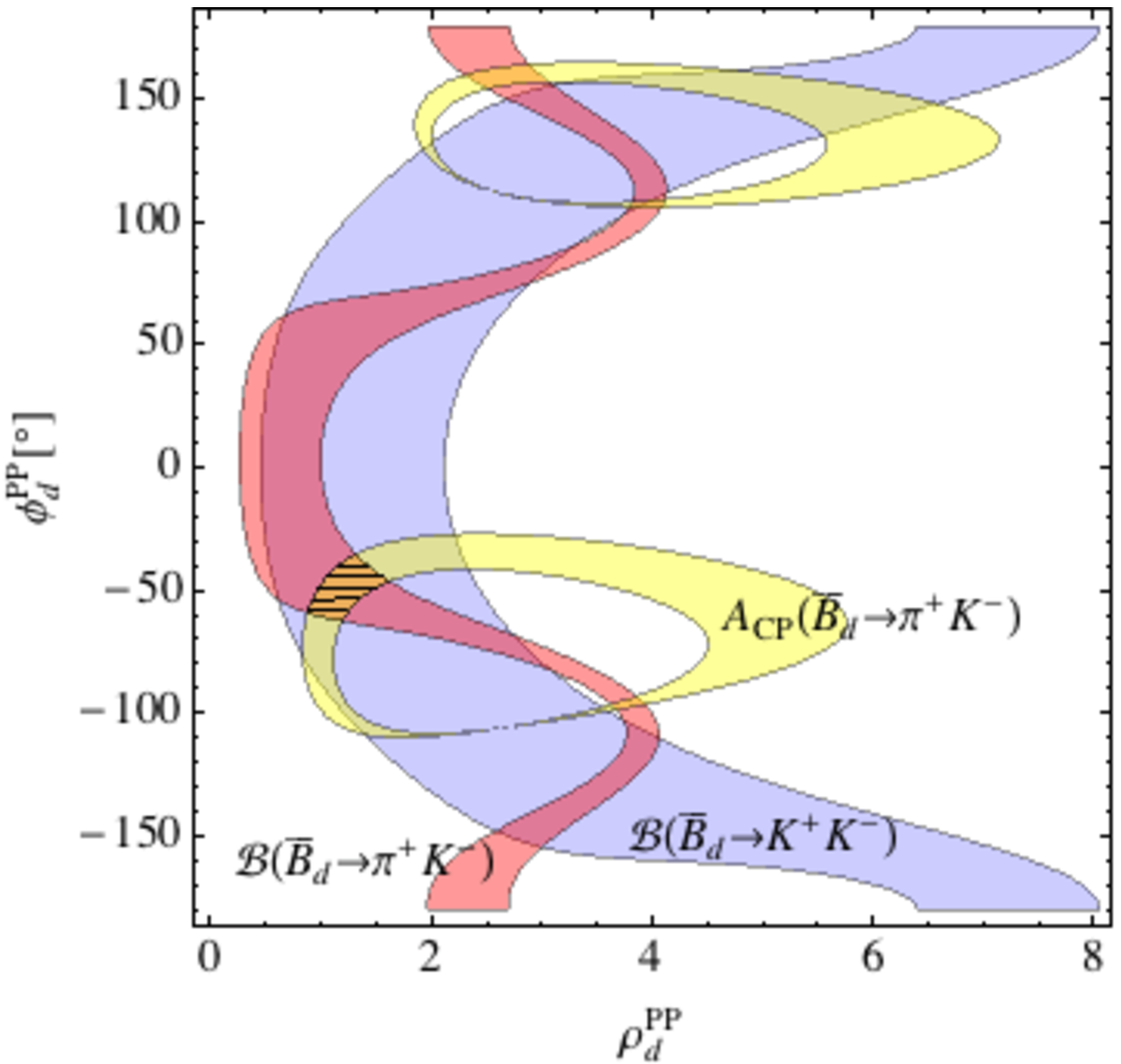}}\\
\subfigure[]{\includegraphics [width=7cm]{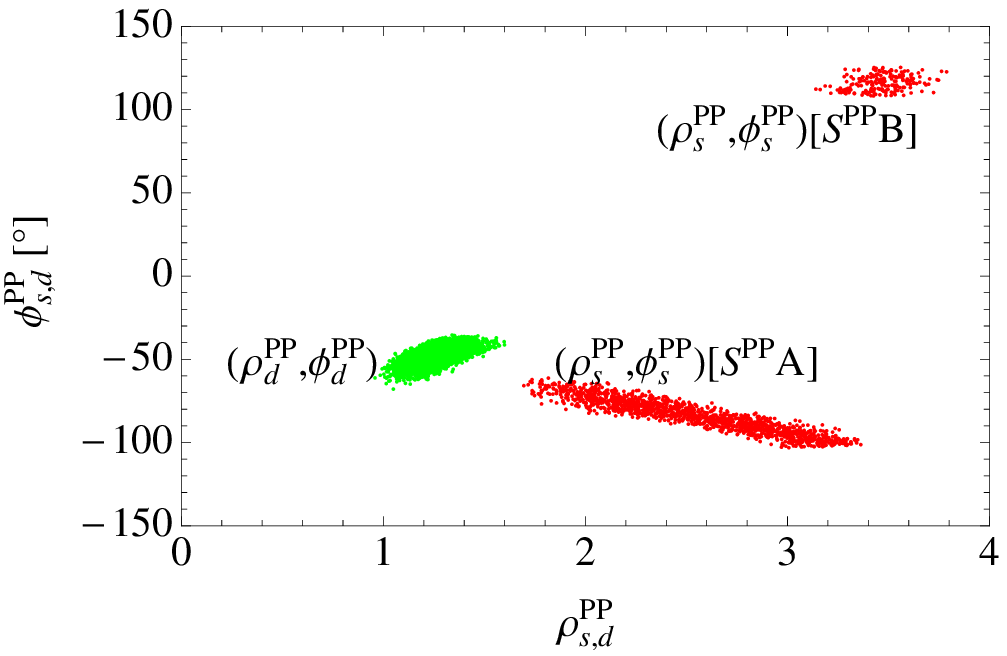}}
\centerline{\parbox{16cm}{\caption{\label{ParaSpac}\small The allowed regions for the annihilation parameters $\phi_{s,d}^{PP}$ and $\rho_{s,d}^{PP}$ under the constraints from the observables labeled in figures, respectively, (Figs.~(a) and (b)) and their combination (Fig.~(c)). }}}
\end{center}
\end{figure}

To keep the predictive power of the QCDF framework, we assume the annihilation parameters are universal for all of the $\bar{B}_s^0\to PP$ decay channels in this paper.  Under the constraints from ${\cal B}(\bar{B}_s^0\to\pi^+\pi^-, \pi^-K^+, K^+K^-)$, $A_{CP}(\bar{B}_s^0\to\pi^-K^+)$ and their combination, the allowed regions for the annihilation parameters $\phi_{s}^{PP}$ and $\rho_{s}^{PP}$ are shown by Figs. \ref{ParaSpac} (a) and (c), respectively. From Fig. \ref{ParaSpac} (a), we find that the traditional treatment $\rho^{PP}\leqslant1$ is allowed by the experimental results of ${\cal B}(\bar{B}_s^0\to\pi^-K^+, K^+K^-)$ and $A_{CP}(\bar{B}_s^0\to\pi^-K^+)$, but obviously excluded by recent experimental measurements ${\cal B}(\bar{B}_s^0\to\pi^+\pi^-)=0.73\pm0.14$. Combining the constraints from above four observables, as Fig.~(c) shows, the annihilation parameters are tightly restricted, and two solutions, named ${\rm S^{PP}A}$ and ${\rm B}$, respectively, are obtained as~\footnote{Out fitting for the annihilation parameters is performed with that the experimental data are allowed within their respectively $1.68\sigma$~($\simeq90\%~\rm{C.L.}$) error bars, while the theoretical uncertainties are also considered by varying the input parameters within their respective regions specified in Appendix B.}
\ba
\left\{\begin{array}{l}
\label{SPPAB}
\rho_s^{PP}=2.5\pm0.8\,,\quad\phi_s^{PP}=-84^{\circ}\pm21^{\circ}\,;\quad {\rm (S^{PP}A)}\\
\rho_s^{PP}=3.5\pm0.3\,,\quad\phi_s^{PP}=116^{\circ}\pm9^{\circ}\,.\quad {\rm (S^{PP}B)}
\end{array}\right.
\ea
Both of them imply a large annihilation correction is essential to accommodate the measured $\bar{B}_s\to PP$ decays.

As a comparison, we  also evaluate the values of the annihilation parameters in $\bar{B}_d^0\to PP$ decays with the constraints from the well measured $\bar{B}_d^0\to \pi^+K^-$ and recent measured $\bar{B}_d^0\to K^+K^-$ decays. From Fig. \ref{ParaSpac} (b), we find  $\phi_{d}^{PP}$ and $\rho_{d}^{PP}$ are bounded strongly by the precise experimental data of the branching fraction and direct CP asymmetry of  $\bar{B}_d^0\to \pi^+K^-$ decay. While, the constraint from ${\cal B}(\bar{B}_d^0\to K^+K^-)$ is weak due to the rough measurement. Corresponding to the allowed region for $\phi_{d}^{PP}$ and $\rho_{d}^{PP}$ shown by Fig. \ref{ParaSpac}~(c)\,, we get the numerical results
\be
\rho_d^{PP}=1.2\pm0.3\,,\quad\phi_d^{PP}=-48^{\circ}\pm16^{\circ}\,,
\ee
which is similar to the result of scenario $\overline{S4}$ given by Eq.~(\ref{S4BdPP}), while significantly different from those by $B_{s}$ decays 
$\phi_{s}^{PP}$ and $\rho_{s}^{PP}$ in Eq.~(\ref{SPPAB}).

For $\bar{B}_s^0\to PV$ decay modes, so far, there is no available experimental measurement  could be used to put a constraint on the annihilation parameters therein. Thus, in our numerical evaluations, we assume that the differences between $\rho(\phi)^{PV,VP}$ and $\rho(\phi)^{PP}$ in $\bar{B}_s^0$ decays are the same as the one in $\bar{B}_d^0$ decays of scenario $\overline{S4}$ given by Eqs.~(\ref{S4BdPP}) and (\ref{S4BdPV}), {\it i.e.},
\ba
\label{SPV}
\rho_s^{PV}=\rho_s^{VP}=\rho_s^{PP}\,,\quad\phi_s^{PV}=\phi_s^{PP}+35^{\circ}\,,\quad\phi_s^{VP}=\phi_s^{PP}-15^{\circ}\,.
\ea

With the default values of $\rho_s^{PP,PV,VP}$ and $\phi_s^{PP,PV,VP}$ given by Eqs.~(\ref{SPPAB})  and (\ref{SPV}) as inputs, we present our  results of the observables in fourth and fifth columns of Table \ref{PPPV}. We find that ${\cal B}(\bar{B}_s^0\to\pi^+\pi^-)$ could be enhanced to $~0.7\times10^{-6}$ to match the experimental data  with large annihilation parameters within QCDF. Furthermore, all of the other theoretical results are in good agreement with the experimental data. The branching fractions of $\bar{B}_s^0\to PV$ decays, which may play an important role to confirm or refute the large annihilation corrections,  are  too small $\sim{\cal O}(10^{-8})$ to be measured very soon at LHCb.  

\subsection{Within Scheme II}
Alternative to the way of the parameterization, the end-point divergency could be regulated by an infrared finite dynamical gluon propagator, which has been successfully applied to the B meson non-leptonic decays~\cite{YYgluon,Chang1,IDGApp}. In this paper we adopt the Cornwall's description for the gluon propagator, which is given by~\cite{Cornwall}
\be
D(q^2)=\frac{1}{q^2-M_g^2(q^2)+i\epsilon}~,
 \label{Dg}
\ee
with the dynamical gluon mass 
\be
M_g^2(q^2)=m_g^2\Bigg[\frac{\mathrm{ln}\Big(\frac{q^2+4m_g^2}{\Lambda_{QCD}^2}\Big)}
{\mathrm{ln}\Big(\frac{4m_g^2}{\Lambda_{QCD}^2}\Big)}\Bigg]^{-\frac{12}{11}},
\label{Mg}
\ee
where $q^2$ is the gluon momentum square, which is space-like in the hard spectator scattering contributions and time-like in the annihilation corrections. The corresponding strong coupling constant reads
\be
\alpha_s(q^2)=\frac{4\pi}{\beta_0\mathrm{ln}\Big(\frac{q^2+4M_g^2(q^2)}{\Lambda_{QCD}^2}\Big)}~,
\label{Alphas}
\ee
where $\beta_0=11-\frac{2}{3}n_f$ is the first coefficient of the beta function, and $n_f$ the number of active flavors. 

With the description given above, the amplitudes of the hard spectator scattering contributions and annihilation corrections in the $B\to PP$ and $PV$ decays have been derived in Ref.~\cite{Chang1}. Within this scheme, it is found that the hard spectator scattering contributions are real and the annihilation corrections are complex with a large imaginary part~\cite{Chang1}. Moreover, the strength of the annihilation correction is sensitive to the sole input parameter, the effective gluon mass scale $m_g$, which typical value is $0.5\pm0.2\,{\rm GeV}$ obtained by relating the gluon mass to the gluon condensate~\cite{Cornwall}. Interestingly, in B meson system, with the constraints from $B_{u,d}\to\pi K\,,\pi K^{\ast}$ and $\rho K$ decays, a reasonable similar result $m_g=0.5\pm0.05\,{\rm GeV}$ is found  in Ref.~\cite{Chang1}.  So, as a crossing check, it is worthy to evaluate the value of the effective gluon mass scale $m_g$ in $\bar{B}_s^0$ decays.

\begin{figure}[t]
\begin{center}
\subfigure[]{\includegraphics [width=5cm]{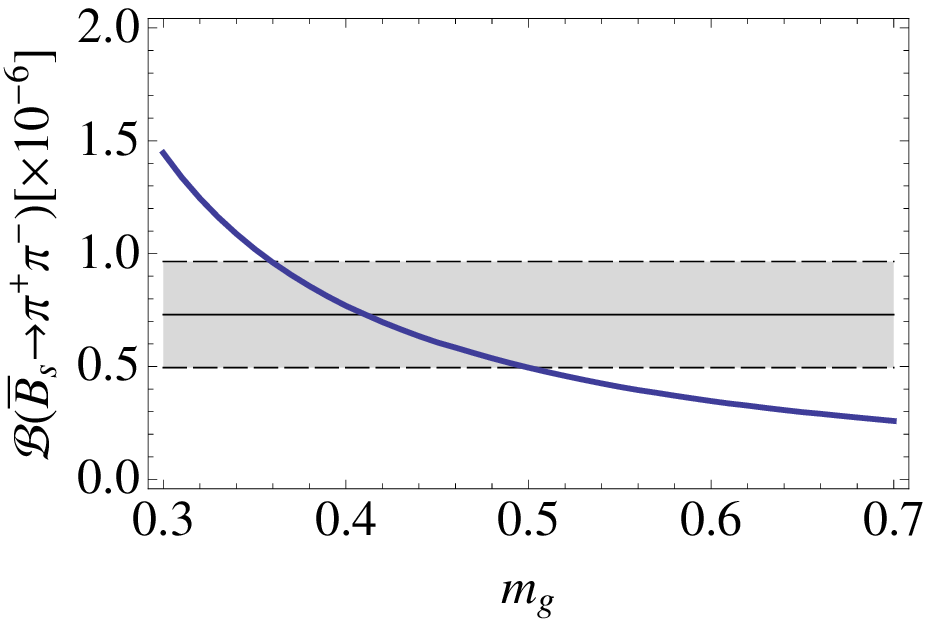}}
~~\subfigure[]{\includegraphics [width=5cm]{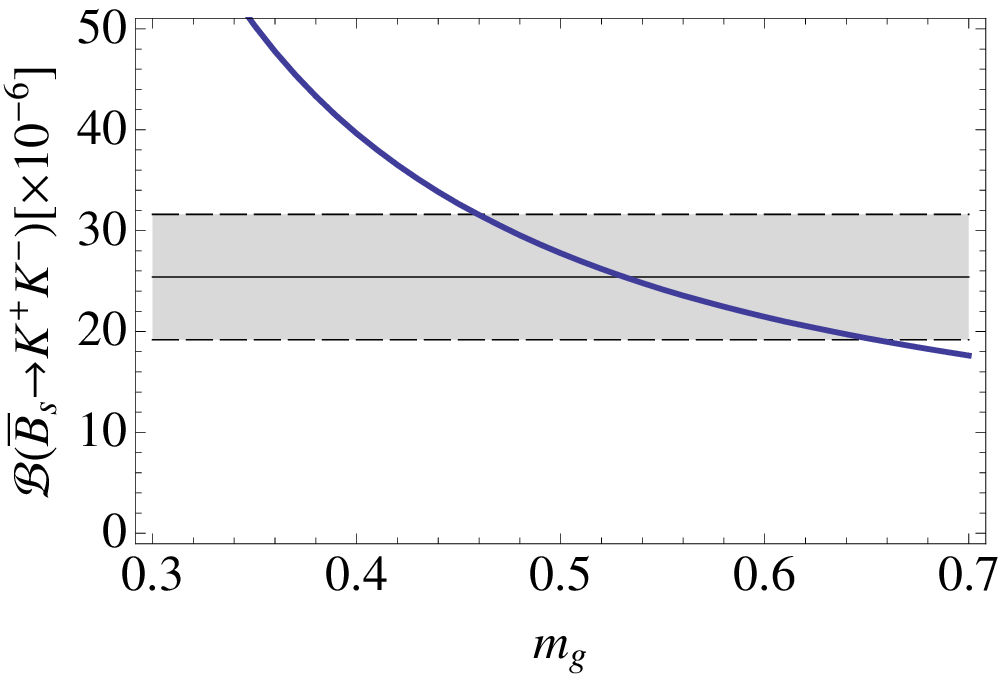}}\\
\subfigure[]{\includegraphics [width=5cm]{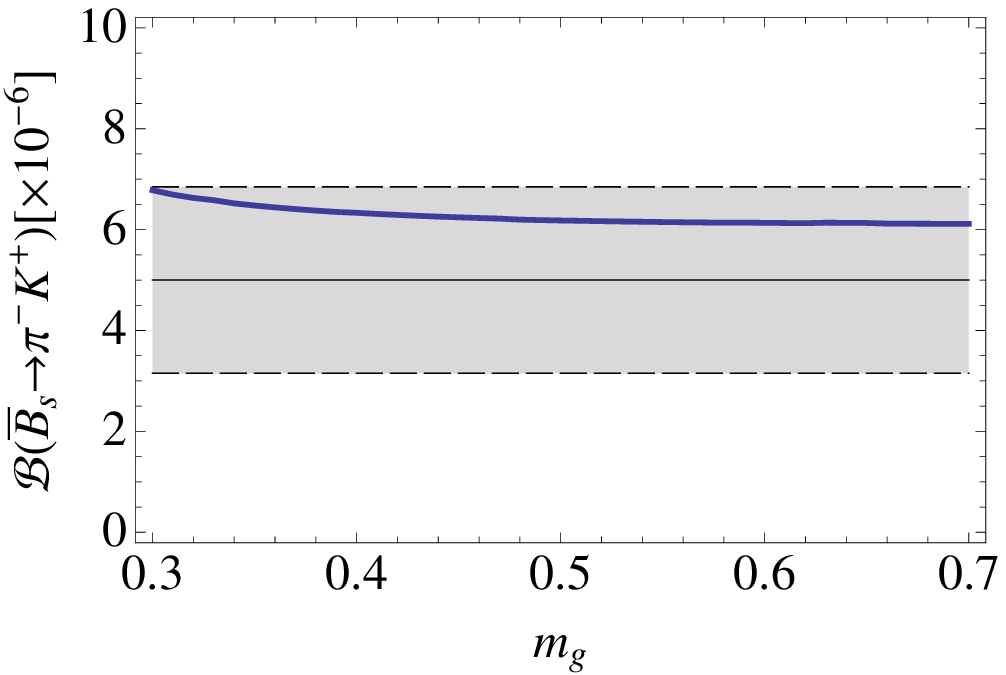}}
~~\subfigure[]{\includegraphics [width=5cm]{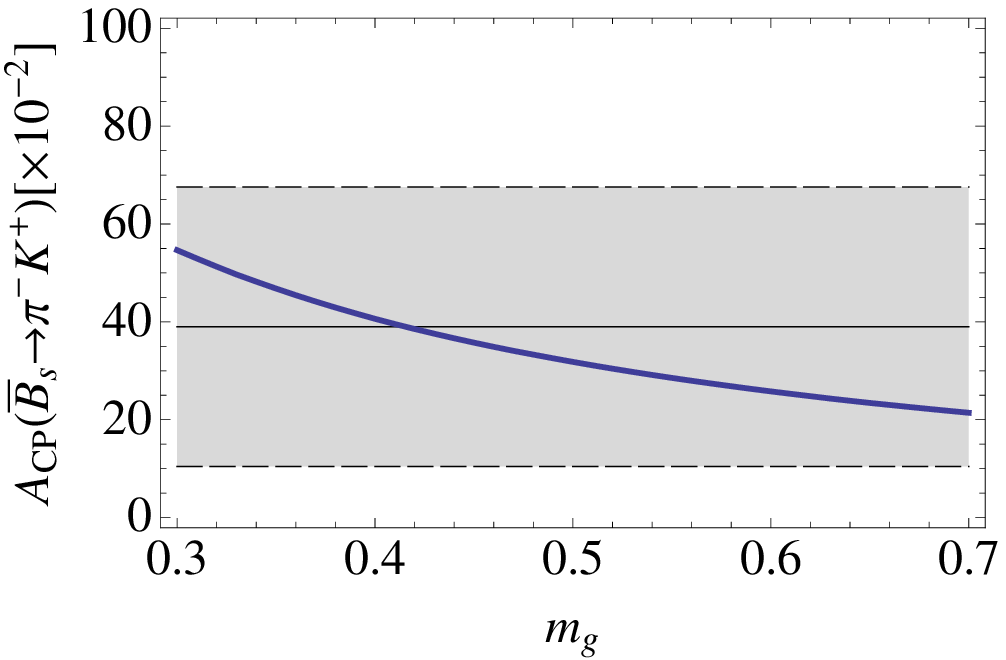}}
\centerline{\parbox{16cm}{\caption{\label{obsgm}\small The dependence of the measured observables of the $\bar{B}_s^0\to PP$ decays on the effective gluon mass scale. The dashed lines correspond to the error bars~($1.68\sigma$).  }}}
\end{center}
\end{figure}

With $m_g=[0.3,0.7]\,{\rm GeV}$ allowed, the dependences of the measured observables ${\cal B}(\bar{B}_s^0\to\pi^+\pi^-\,,K^+K^-\,,\pi^-K^+)$ and $A_{CP}(\bar{B}_s^0\to\pi^-K^+)$ on $m_g$ are shown in Fig.~\ref{obsgm}. From Fig.~\ref{obsgm}~(a), we find that a small $m_g\sim0.43\,{\rm GeV}$, which would lead to large annihilation corrections, is required by large experimental data ${\cal B}(\bar{B}_s^0\to\pi^+\pi^-)=0.73\pm0.14$. While, as Fig.~\ref{obsgm}~(b) shows, a relative large $m_g\sim0.52\,{\rm GeV}$ could result in a good agreement between the theoretical prediction and experimental data for ${\cal B}(\bar{B}_s^0\to K^+K^-)$. With the experimental errors considered, one also could find a common solution 
\be
m_g=0.48\pm0.02\,{\rm GeV}\,.
\ee
Where, its upper limit is dominated by ${\cal B}(\bar{B}_s^0\to\pi^+\pi^-)$, and the lower one is dominated by ${\cal B}(\bar{B}_s^0\to K^+K^-)$. Moreover,  due to that a larger annihilation strength  is required by ${\cal B}(\bar{B}_s^0\to\pi^+\pi^-)$, such a solution is a bit smaller than the finding $m_g=0.5\pm0.05\,{\rm GeV}$ in $B_{u,d}$ decays~\cite{Chang1},  although they are also in agreement. Due to the dominance of the tree contribution $\alpha_1$ in the amplitude of $\bar{B}_s^0\to\pi^+K^-$ decay, the effect of the annihilation contributions related to $m_g$ to ${\cal B}(\bar{B}_s^0\to\pi^+K^-)$ is negligible, which can be seen from  Fig.~\ref{obsgm}~(c). Furthermore, as Fig.~\ref{obsgm}~(d) shows, because of the large experimental error, the constraint from $A_{CP}(\bar{B}_s^0\to\pi^+K^-)$ on $m_g$ is weak too.

Taking $m_g=0.48\,{\rm GeV}$, our numerical results for the observables are listed in the sixth column of Table ~\ref{PPPV}. One may find all of the results are in good agreement with the experimental data, and most of them are similar to the scenarios ${\rm S^{PP}A}$ and ${\rm S^{PP}B}$ in scheme~I. Within scheme~I,  $A_{CP}(\bar{B}_s^0\to\pi^+\rho^-)$ and  $A_{CP}(\bar{B}_s^0\to\pi^-\rho^+)$ could be large due to the assumption that $\phi_s^{PV}\neq\phi_s^{VP}$, which can be seen from Eq.~(\ref{SPV}). However, within scheme~II, we find that $A_{CP}(\bar{B}_s^0\to\pi^+\rho^-)=A_{CP}(\bar{B}_s^0\to\pi^-\rho^+)\approx0$ with any value of $m_g$. The significantly different predictions for such two observables within two schemes will possibly be judged by the running LHCb and upcoming super-B experiments.

\section{$\bar{B}_s^0\to VV$ decay modes}
Compared with $B\to PP$ and $PV$ decays, the $B\to VV$ decays involve more observables, which are sensitive probes  for testing the SM and various calculation approaches. Recently,  $\bar{B}_s^0\to K^{\ast 0}\bar{K}^{\ast 0}$ and $\phi\phi$ decays have been measured by LHCb and CDF collaborations, respectively, 
\ba\label{LHCbKstar}
\left\{\begin{array}{l}
{\cal B}(\bar{B}_s^0\to K^{\ast 0}\bar{K}^{\ast 0})=(28.1\pm4.6({\rm stat.})\pm4.5({\rm syst.})\pm3.4(f_s/f_d))\times 10^{-6}\,,\\
f_L(\bar{B}_s^0\to K^{\ast 0}\bar{K}^{\ast 0})=0.31\pm0.12({\rm stat.})\pm0.04({\rm syst.})\,,\qquad{\rm LHCb}~\cite{LHCbKstar}\\
f_{\bot}(\bar{B}_s^0\to K^{\ast 0}\bar{K}^{\ast 0})=0.38\pm0.11({\rm stat.})\pm0.04({\rm syst.})\,;
\end{array}\right.
\ea
and 
\ba\label{CDFphi}
\left\{\begin{array}{l}
{\cal B}(\bar{B}_s^0\to\phi\phi) =(23.2\pm1.8({\rm stat.})\pm8.2({\rm syst.}))\times 10^{-6}\,,\\
f_L(\bar{B}_s^0\to\phi\phi)=0.348\pm0.041({\rm stat.})\pm0.021({\rm syst.})\,,\qquad{\rm CDF}~\cite{CDFphi}\\
f_{\bot}(\bar{B}_s^0\to\phi\phi)=0.365\pm0.044({\rm stat.})\pm0.027({\rm syst.})\,.
\end{array}\right.
\ea
Because of the left-handedness of the weak interaction and the fact that the high-energy QCD interactions conserve helicity, the SM expects the dominance of the longitudinal amplitude and  the transverse  amplitudes are suppressed by the factor $m_{\phi\,,\bar{K}^{\ast 0}}/m_B$, which significantly conflicts with the LHCb and CDF observation $f_L(\bar{B}_s^0\to K^{\ast 0}\bar{K}^{\ast 0}\,,\phi\phi)\sim f_{\bot}(\bar{B}_s^0\to K^{\ast 0}\bar{K}^{\ast 0}\,,\phi\phi)$. Therefore, it is worthy to perform a detailed evaluation within QCDF, and check if the tension could be moderated by annihilation corrections. 
 
\begin{table}[t]
 \begin{center}
 \caption{The numerical results for the branching fractions ${\cal B} [\times10^{-6}]$, the direct CP violations $A_{CP}[\times10^{-2}]$, longitudinal and  transverse polarization fractions $f_{L,\bot}\,[\times10^{-2}]$ for $\bar{B}_s^0\to\rho\rho$, $K^{\ast 0}\bar{K}^{\ast 0}$ and $\phi\phi$ decays in each scenarios.}
 \label{VV}
 \vspace{0.5cm}
 \small
 \doublerulesep 0.1pt \tabcolsep 0.05in
 \begin{tabular}{lccccccccccc} \hline \hline
                                                                  &Exp                                                      &\multicolumn{5}{c}{Scheme I}                                                                                                                                  &\multicolumn{1}{c}{Scheme II}\\
                                                                 &                           & $\overline{S4}$                               & $\rm S^{VV}A$                        & $\rm S^{VV}B$              & $\rm S^{VV}C$              & $\rm S^{VV}D$          &$m_g=0.50{\rm GeV}$     \\ \hline
 ${\cal B}(\bar{B}_s^0\to\rho^+\rho^-)$&---                   &$0.36^{+0.12}_{-0.09}$      &$0.24^{+0.12}_{-0.09}$     &$0.66^{+0.22}_{-0.18}$   &$0.67^{+0.23}_{-0.18}$ &$0.21^{+0.10}_{-0.08}$ &$1.30^{+0.44}_{-0.34}$\\
 ${\cal B}(\bar{B}_s^0\to\rho^0\rho^0)$&---                   &$0.18^{+0.06}_{-0.05}$      &$0.12^{+0.06}_{-0.04}$     &$0.33^{+0.11}_{-0.09}$  &$0.33^{+0.11}_{-0.09}$ &$0.10^{+0.05}_{-0.04}$ &$0.65^{+0.22}_{-0.17}$\\
 $A_{CP}(\bar{B}_s^0\to\rho^+\rho^-)$&---                   &$0$                                        &$0$                                           &$0$                                   &$0$                                   &$0$                  &$0$\\
 $A_{CP}(\bar{B}_s^0\to\rho^0\rho^0)$&---                  &$0$                                      &$0$                                       &$0$                                      &$0$                                   &$0$                     &$0$\\
  $f_L(\bar{B}_s^0\to\rho^+\rho^-)$       &---                       &$99^{+0}_{-0}$            &$96^{+0}_{-1}$                   &$98^{+0}_{-0}$                 &$98^{+0}_{-0}$              &$97^{+1}_{-1}$   &$98^{+0}_{-0}$ \\
 $f_L(\bar{B}_s^0\to\rho^0\rho^0)$       &---                        &$99^{+0}_{-0}$            &$96^{+0}_{-1}$                   &$98^{+0}_{-0}$                  &$98^{+0}_{-0}$             &$97^{+1}_{-1}$    &$98^{+0}_{-0}$ \\
  \hline  \hline
${\cal B}(\bar{B}_s^0\to K^{\ast 0}
\bar{K}^{\ast 0})$                                   &$28.1\pm6.5$       &$11.0^{+2.9}_{-2.6}$     &$16.7^{+6.5}_{-5.0}$     &$15.3^{+4.9}_{-3.7}$         &$15.9^{+5.4}_{-3.7}$     &$15.9^{+6.5}_{-4.9}$         &$20.6^{+6.5}_{-5.2}$\\
${\cal B}(\bar{B}_s^0\to\phi\phi)$        &$23.2\pm8.4$       &$21.9^{+10.6}_{-4.6}$    &$41.6^{+18.8}_{-12.0}$   &$39.7^{+19.0}_{-10.0}$    &$38.0^{+15.7}_{-10}$    &$41.9^{+19.4}_{-12.0}$    &$49.9^{+25.6}_{-13.3}$\\
$A_{CP}(\bar{B}_s^0\to K^{*0}
\bar{K}^{*0})$                                      &---                         &$0.8^{+0.1}_{-0.1}$         &$-0.1^{+0.1}_{-0.0}$      &$0.6^{+0.1}_{-0.1}$           &$0.1^{+0.2}_{-0.1}$      &$0.5^{+0.1}_{-0.1}$        &$0.5^{+0.1}_{-0.1}$    \\
$A_{CP}(\bar{B}_s^0\to\phi\phi)$       &---                         &$0.9^{+0.2}_{-0.1}$        &$-0.1^{+0.2}_{-0.0}$       &$0.6^{+0.3}_{-0.1}$           &$-0.0^{+0.2}_{-0.1}$     &$0.5^{+0.3}_{-0.1}$      &$0.5^{+0.2}_{-0.1}$\\
 $f_L(\bar{B}_s^0\to K^{\ast 0}
 \bar{K}^{\ast 0})$                                 &$31\pm13$          &$71^{+7}_{-7}$              &$41^{+4}_{-3}$                &$42^{+10}_{-6}$                 &$45^{+10}_{-7}$          &$38^{+3.3}_{-1.8}$      &$65^{+7}_{-6}$\\
$f_L(\bar{B}_s^0\to\phi\phi)$                &$34.8\pm4.6$       &$56^{+11}_{-8}$           &$36^{+4}_{-3}$              &$34^{+7}_{-3}$                   &$32^{+8}_{-2}$              &$35^{+4}_{-3}$           &$57^{+9}_{-4}$\\
$f_\bot(\bar{B}_s^0\to K^{\ast 0}
\bar{K}^{\ast 0})$                                &$38\pm12$            &$13^{+4}_{-4}$              &$27^{+2}_{-2}$              &$26^{+3}_{-5}$                    &$24^{+4}_{-4}$             &$29^{+2}_{-2}$                &$15^{+3}_{-3}$\\
$f_\bot(\bar{B}_s^0\to\phi\phi)$            &$36.5\pm5.2$       &$21^{+4}_{-5}$             &$31^{+2}_{-2}$               &$31^{+2}_{-4}$                   &$32^{+2}_{-4}$              &$32^{+2}_{-2}$         &$21^{+2}_{-4}$\\
 \hline \hline
 \end{tabular}
 \end{center}
 \end{table}

\subsection{Within Scheme I}
With the annihilation parameters given by Eq. (\ref{S4VV}), the prediction of  scenarios $\overline{S4}$ for $\bar{B}_s^0\to\rho\rho\,,K^{\ast 0}\bar{K}^{\ast 0}$ and $\phi\phi$ decays are listed in the third column of Table \ref{VV}, which agree with the former results of the QCDF~\cite{Beneke3,Cheng2}. One may find that there are no significant direct CP asymmetries for these decay modes, and the longitudinal polarization fractions of $\bar{B}_s^0\to\rho\rho$ decays are close to unity. The branching fraction of the $\bar{B}_s^0\to \phi\phi$ decay agrees well with the experimental data. While, the default result ${\cal B}(\bar{B}_s^0\to K^{\ast 0}\bar{K}^{\ast 0})\sim11.0\times10^{-6}$ is significantly smaller than LHCb measurement $\sim28.1\times10^{-6}$. However, one may notice that the uncertainties in the experimental measurement are still very large. For their polarization fractions, as expected above, the prediction of  scenarios $\overline{S4}$ implies $f_L(\bar{B}_s^0\to K^{\ast 0}\bar{K}^{\ast 0}\,,\phi\phi)\sim 0.71\,,0.56>f_\bot(\bar{B}_s^0\to K^{\ast 0}\bar{K}^{\ast 0}\,,\phi\phi)\sim0.13\,,0.21$, which conflict with the LHCb and CDF observation $f_L(\bar{B}_s^0\to K^{\ast 0}\bar{K}^{\ast 0}\,,\phi\phi)\approx f_{\bot}(\bar{B}_s^0\to K^{\ast 0}\bar{K}^{\ast 0}\,,\phi\phi)$. In the following, we would perform a detailed evaluations to check whether such a discrepancy could be moderated  by the annihilation corrections. 

\begin{figure}[t]
\begin{center}
\subfigure[]{\includegraphics [width=7cm]{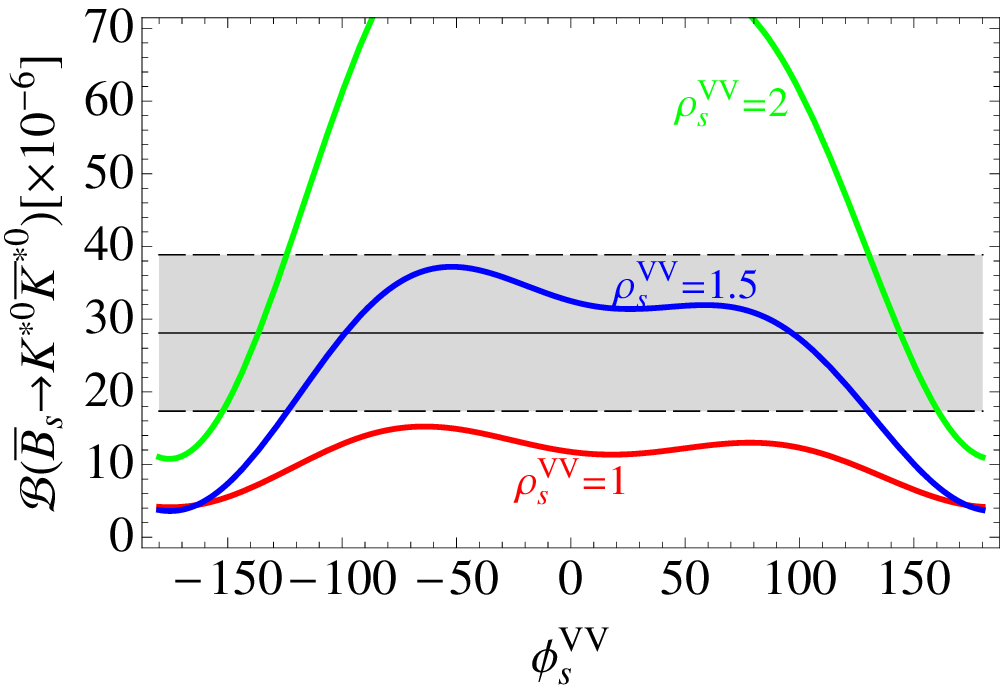}}
~~~~~~~~~~\subfigure[]{\includegraphics [width=7cm]{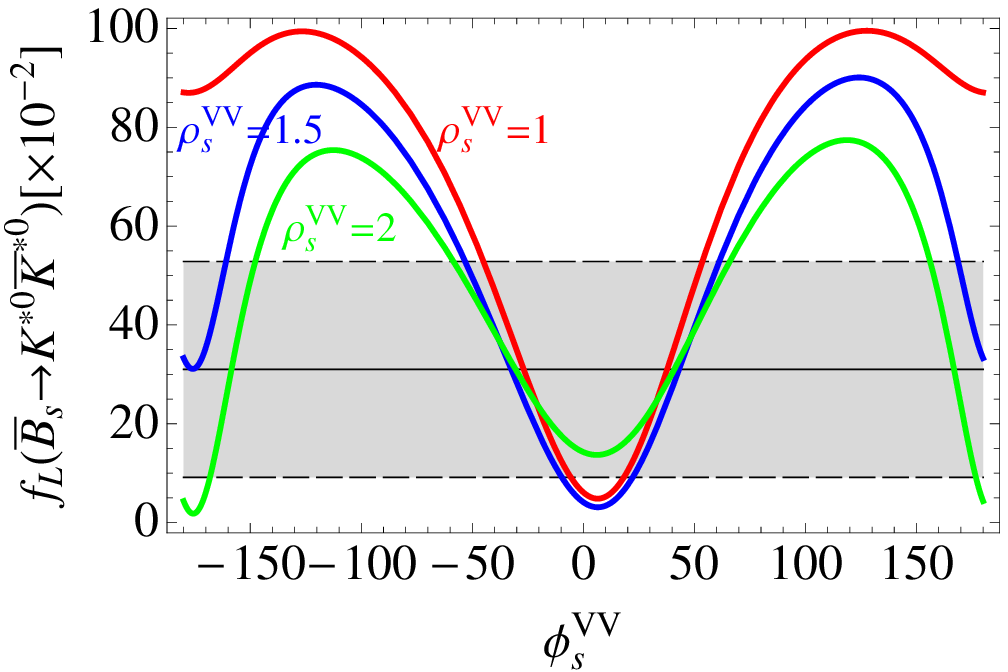}}\\
\subfigure[]{\includegraphics [width=7cm]{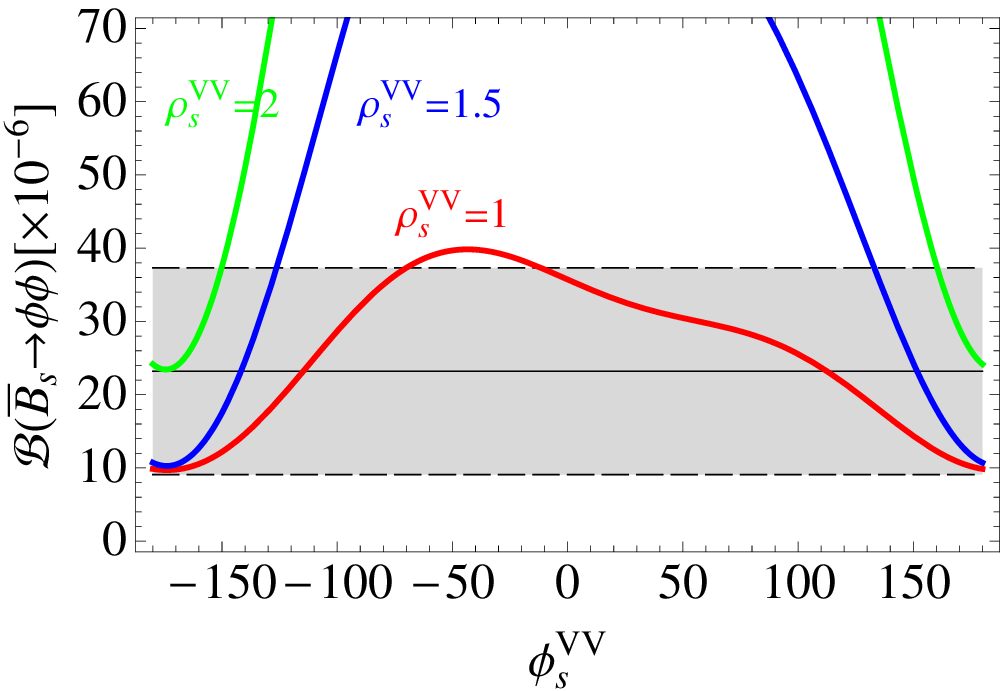}}
~~~~~~~~~~\subfigure[]{\includegraphics [width=7cm]{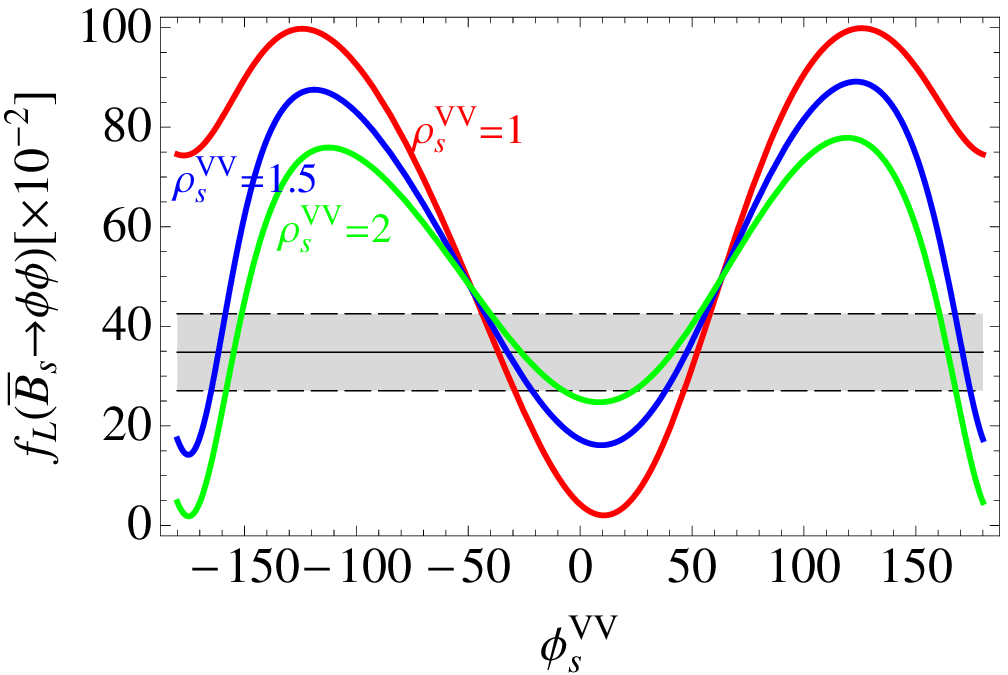}}\\
\centerline{\parbox{16cm}{\caption{\label{obsVVpara}\small The dependence of ${\cal B}(\bar{B}_s^0\to K^{\ast 0}\bar{K}^{\ast 0}\,,\phi\phi)$ and  $f_L(\bar{B}_s^0\to K^{\ast 0}\bar{K}^{\ast 0}\,,\phi\phi)$ on the phases $\phi_{s}^{VV}$ with different $\rho_{s}^{VV}$ values. The dashed lines correspond to the error bars~($1.68\sigma$). }}}
\end{center}
\end{figure}

The dependence of ${\cal B}(\bar{B}_s^0\to K^{\ast 0}\bar{K}^{\ast 0}\,,\phi\phi)$ and  $f_L(\bar{B}_s^0\to K^{\ast 0}\bar{K}^{\ast 0}\,,\phi\phi)$ on the annihilation parameters is shown by Fig.~\ref{obsVVpara}. Comparing Fig.~\ref{obsVVpara} (b) with (d), we find the constraint from $f_L(\bar{B}_s^0\to K^{\ast 0}\bar{K}^{\ast 0})$ on the annihilation parameters is weak due to its large experimental uncertainties. From Fig.~\ref{obsVVpara} (d), one may find the phase $\phi_{s}^{VV}\sim-40^{\circ}$ or $50^{\circ}$ with any value of $\rho_{s}^{VV}$ could be  helpful  to moderate the tension between the theoretical prediction and experimental result for $f_L(\bar{B}_s^0\to\phi\phi)$. While, with such $\phi_{s}^{VV}$ value, as  shown in Figs~\ref{obsVVpara} (a) and (c), a small $\rho_{s}^{VV}\sim1$ is required by both ${\cal B}(\bar{B}_s^0\to K^{\ast 0}\bar{K}^{\ast 0})$ and ${\cal B}(\bar{B}_s^0\to\phi\phi)$. Furthermore, compared with such solutions, we also find a larger  $\rho_{s}^{VV}\sim2$ with a larger phase $\phi_{s}^{VV}\sim-150^{\circ}$ or $160^{\circ}$ are also possible solutions. A detailed  numerical examination is performed in the following in due.

\begin{figure}[t]
\begin{center}
\subfigure[]{\includegraphics [width=6cm]{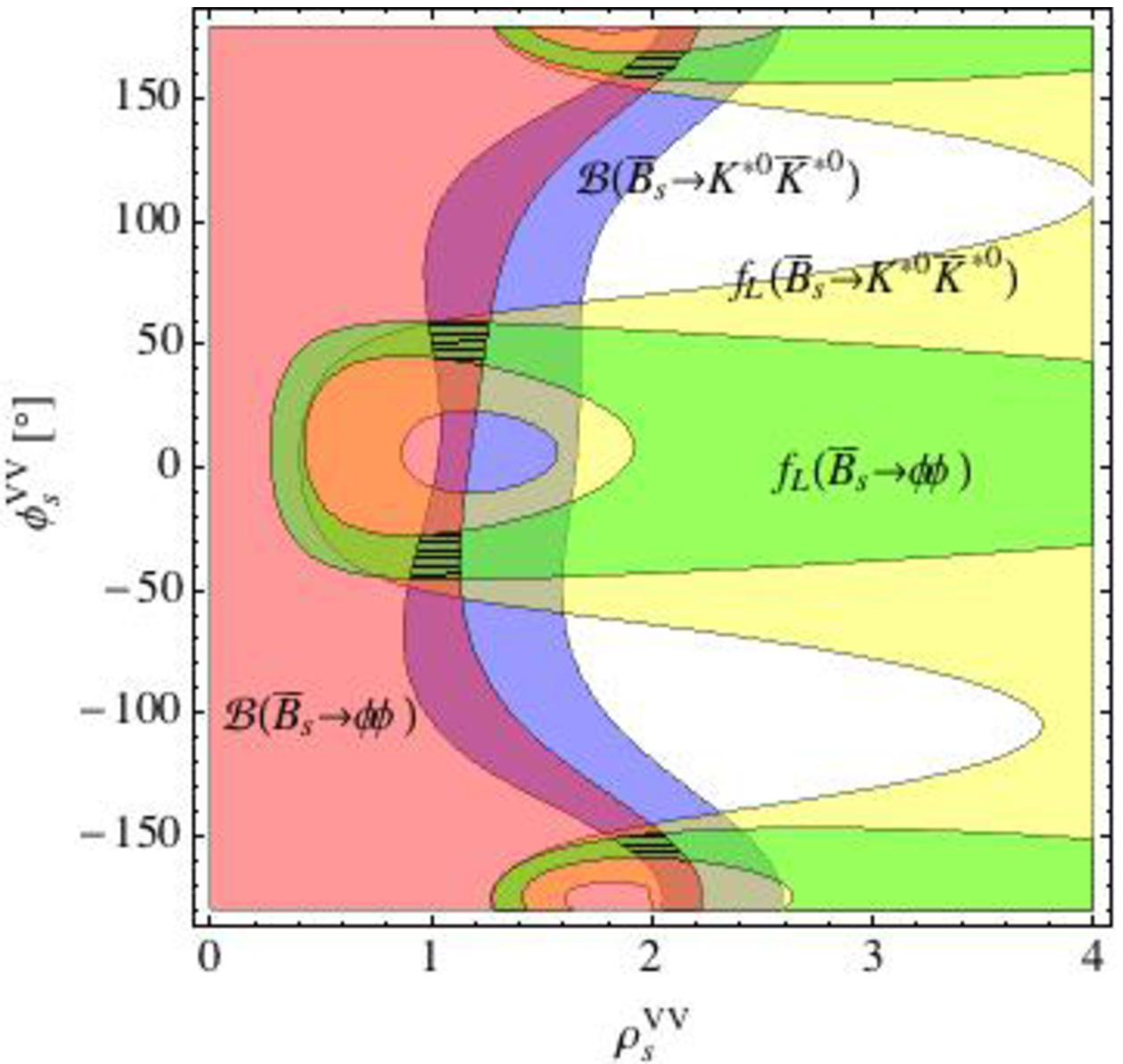}}
~~~~~~~~~~\subfigure[]{\includegraphics [width=7.7cm]{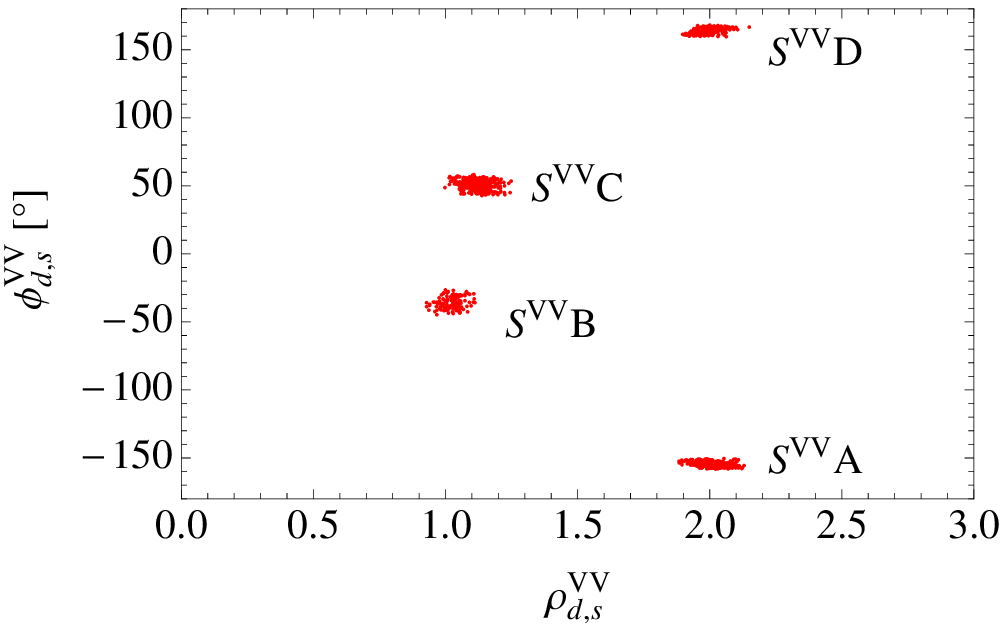}}
\centerline{\parbox{16cm}{\caption{\label{ParaVVSpac}\small The allowed regions for the annihilation parameters $\phi_{s}^{VV}$ and $\rho_{s}^{VV}$ under the constraints from ${\cal B}(\bar{B}_s^0\to K^{\ast 0}\bar{K}^{\ast 0}\,,\phi\phi)$, $f_L(\bar{B}_s^0\to K^{\ast 0}\bar{K}^{\ast 0}\,,\phi\phi)$~(Fig.~(a)) and their combination ~(Fig.~(b)), respectively. }}}
\end{center}
\end{figure}

Under the constraints from ${\cal B}(\bar{B}_s^0\to K^{\ast 0}\bar{K}^{\ast 0}\,,\phi\phi)$ and $f_L(\bar{B}_s^0\to K^{\ast 0}\bar{K}^{\ast 0}\,,\phi\phi)$, the allowed regions for the annihilation parameters are shown in Fig.~\ref{ParaVVSpac}. As shown in Fig.~\ref{ParaVVSpac}~(a), the space of the annihilation parameters are strictly restricted. The upper and the lower limits of $\rho_{s}^{VV}$ are dominated by ${\cal B}(\bar{B}_s^0\to\phi\phi)$ and ${\cal B}(\bar{B}_s^0\to K^{\ast 0}\bar{K}^{\ast 0})$, respectively. While, the ranges of $\phi_{s}^{VV}$ are dominated by $f_L(\bar{B}_s^0\to\phi\phi)$. Finally, under their combined constraints, we could find four solutions shown by Fig.~\ref{ParaVVSpac}~(b). The corresponding numerical results are
\ba\label{SVV}
\left\{\begin{array}{l}
\rho_s^{VV}=2.0\pm0.1\,,\quad\phi_s^{VV}=-154^{\circ}\pm4^{\circ}\quad {\rm (S^{VV}A)},\\
\rho_s^{VV}=1.0\pm0.1\,,\quad\phi_s^{VV}=-36^{\circ}\pm9^{\circ}\quad {\rm (S^{VV}B)},\\
\rho_s^{VV}=1.1\pm0.1\,,\quad\phi_s^{VV}=50^{\circ}\pm7^{\circ}\quad {\rm (S^{VV}C)},\\
\rho_s^{VV}=2.0\pm0.1\,,\quad\phi_s^{VV}=164^{\circ}\pm4^{\circ}\quad {\rm (S^{VV}D)}.
\end{array}\right.
\ea

With the default values of the annihilation parameters given by Eq.~(\ref{SVV}),  our predictions of scheme~I are summarized  in the Table~\ref{VV}.  We find that $f_L(\bar{B}_s^0\to K^{\ast 0}\bar{K}^{\ast 0}\,,\phi\phi)$  could reduced to the experimental data by the annihilation contributions.  Meanwhile, $f_{\bot}(\bar{B}_s^0\to K^{\ast 0}\bar{K}^{\ast 0}\,,\phi\phi)$ are also significantly enhanced, and agree well with the experimental data. However, similar to the case in  scenarios $\overline{S4}$, the  result of ${\cal B}(\bar{B}_s^0\to\phi\phi)$ is larger than the one of ${\cal B}(\bar{B}_s^0\to K^{\ast 0}\bar{K}^{\ast 0})$ by a factor about 2. In the previous works, the theoretical predictions within both QCDF and pQCD frameworks, for example 
\ba\label{QCDF1}
&&\left\{\begin{array}{l}
{\cal B}(\bar{B}_s^0\to\phi\phi)=(21.8^{+1.1+30.4 }_{-1.1-17.0})\times 10^{-6}\,,\\
{\cal B}(\bar{B}_s^0\to K^{\ast 0}\bar{K}^{\ast 0})=(9.1^{+0.5+11.3}_{-0.4-6.8})\times 10^{-6}\,;
\end{array}\qquad{\rm QCDF}~\cite{Beneke3}\right.\\
\label{QCDF2}
&&\left\{\begin{array}{l}
{\cal B}(\bar{B}_s^0\to\phi\phi)=(16.7^{+2.6+11.3}_{-2.1-8.8})\times 10^{-6}\,,\\
{\cal B}(\bar{B}_s^0\to K^{\ast 0}\bar{K}^{\ast 0})=(6.6^{+1.1+1.9 }_{-1.4-1.7})\times 10^{-6}\,;
\end{array}\qquad{\rm QCDF}~\cite{Cheng2}\right.\\
\label{pQCD}
&&\left\{\begin{array}{l}
{\cal B}(\bar{B}_s^0\to\phi\phi)=(35.3^{+8.3+16.7+0.0 }_{-6.9-10.2-0.0})\times 10^{-6}\,,\\
{\cal B}(\bar{B}_s^0\to K^{\ast 0}\bar{K}^{\ast 0})=(7.8^{+1.9+3.8+0.0}_{ -1.5-2.2-0.0})\times 10^{-6}\,,
\end{array}\qquad{\rm pQCD}~\cite{relaRef}\right.
\ea
present a similar result that ${\cal B}(\bar{B}_s^0\to\phi\phi)>{\cal B}(\bar{B}_s^0\to K^{\ast 0}\bar{K}^{\ast 0})$, which is obviously inconsistent with  the LHCb and CDF measurements Eqs.~(\ref{LHCbKstar}) and (\ref{CDFphi}) that ${\cal B}(\bar{B}_s^0\to\phi\phi)\simeq{\cal B}(\bar{B}_s^0\to K^{\ast 0}\bar{K}^{\ast 0})$.  
Such a theoretical situation could be easily understood from their amplitudes given by Eqs.~(\ref{amp5_SM}) and (\ref{amp6_SM}) in appendix A. The amplitudes of both $\bar{B}_s^0\to K^{\ast 0}\bar{K}^{\ast 0}$ and $\bar{B}_s^0\to\phi\phi$ decays are dominated by the effective coefficients $\alpha_4^{p}$, and annihilation contributions to them are similar. However, an additional overall factor 2 is included in the amplitude of $\bar{B}_s^0\to\phi\phi$. So,  ${\cal B}(\bar{B}_s^0\to\phi\phi)$ would be always larger than ${\cal B}(\bar{B}_s^0\to K^{\ast 0}\bar{K}^{\ast 0})$.

For the $\bar{B}_s^0\to\rho\rho$ decays, their longitudinal polarization fractions are always close to unity within the four solutions. While, the predictions of $S^{VV}B$ and $S^{VV}C$ for ${\cal B}(\bar{B}_s^0\to\rho\rho)$ are significantly larger than the ones of $S^{VV}A$ and $S^{VV}D$. So, the four solutions for the annihilation parameters given by Eq.~(\ref{SVV}) could be tested by the upcoming LHC-b measurements of   ${\cal B}(\bar{B}_s^0\to\rho\rho)$.

\subsection{Within Scheme II}
With the infrared finite gluon propagator to deal with the endpoint divergences, the hard spectator and the annihilation corrections for $B\to PP$ and $PV$ decays have been evaluated in Ref.~\cite{Chang1}. While, the ones for $B\to VV$ decays have not calculated until now. So, firstly, we would re-calculate these corrections within scheme~II. With the same convention and notation as Refs.~\cite{Chang1} and~\cite{Beneke3}, the hard spectator scattering contributions can be expressed as
\begin{eqnarray}
H_i^-=-\frac{2f_Bf_{V_1}^\bot}{m_Bm_bF_-^{B\to V_1}(0)}\int_{0}^{1}dxdyd\xi\alpha_s(q^2)\frac{\Phi_{B1}\phi_1^\bot(x)\phi_{b2}(y)}{(\xi\bar{x}+\omega^2(q^2))\bar{x}y}\,,
\end{eqnarray}
for $i=1,2,3,4,9,10$;
\begin{eqnarray}
H_i^-=\frac{2f_Bf_{V_1}^\bot}{m_Bm_bF_-^{B\to V_1}(0)}\int_{0}^{1}dxdyd\xi\alpha_s(q^2)\frac{\Phi_{B1}\phi_1^\bot(x)\phi_{a2}(y)}{(\xi\bar{x}+\omega^2(q^2))\bar{x}\bar{y}}\,,
\end{eqnarray}
for $i=5,7$;
\begin{eqnarray}
H_i^-=-\frac{f_Bf_{V_1}}{m_BF_-^{B\to V_1}(0)}\frac{m_1}{m_2^2}\int_{0}^{1}dxdyd\xi\alpha_s(q^2)\frac{\Phi_{B1}\phi_{a1}(x)\phi_2^\bot(y)}{(\xi\bar{x}+\omega^2(q^2))\bar{y}y}
\end{eqnarray}
for $i=6,8$. In which, $\omega^2(q^2)=M_g^2(q^{2})/M_B^2$, $q^{2}=-Q^{2}$ and $Q^2\simeq-\xi\bar{x}M_B^2$ is the space-like gluon momentum square in the scattering kernels. The function $\Phi_{B1}(\xi)$ is the B meson light-cone distribution amplitude, where $\xi$ is the light-cone momentum fraction of the light anti-quark in the B meson. In our following numerical evaluation, $\Phi_B(\xi)=N_B\xi(1-\xi)\textmd{exp}\Big[-\Big(\frac{M_B}{M_B-m_b}\Big)^2(\xi-\xi_B)^2\Big]$~\cite{YYPhiB} is used. 

The longitudinal part of the annihilation amplitudes are given by
\begin{eqnarray}
A_1^{i,0}&=&\pi{\int}_{0}^{1}dxdy\alpha_s(q^2)\Big\{\Phi_{V1}(x)\Phi_{V2}(y)\big[\frac{\bar{y}}{(x\bar{y}-\omega^2(q^2)+i\varepsilon)(1-\bar{x}y)}+\frac{1}{\bar{y}(x\bar{y}-\omega^2(q^2)+i\varepsilon)}\big]\nonumber\\
&&-r_\chi^{V_1}r_\chi^{V_2}\Phi_{v1}(x)\Phi_{v2}(y)\frac{2}{x\bar{y}-\omega^2(q^2)+i\varepsilon}\Big\}\,,\\
A_2^{i,0}&=&\pi{\int}_{0}^{1}dxdy\alpha_s(q^2)\Big\{\Phi_{V1}(x)\Phi_{V2}(y)\big[\frac{x}{(x\bar{y}-\omega^2(q^2)+i\varepsilon)(1-\bar{x}y)}+\frac{1}{x(x\bar{y}-\omega^2(q^2)+i\varepsilon)}\big]\nonumber\\
&&-r_\chi^{V_1}r_\chi^{V_2}\Phi_{v1}(x)\Phi_{v2}(y)\frac{2}{x\bar{y}-\omega^2(q^2)+i\varepsilon}\Big\}\,,\\
A_3^{i,0}&=&\pi{\int}_{0}^{1}dxdy\alpha_s(q^2)\Big\{r_\chi^{V_1}\Phi_{v1}(x)\Phi_{V2}(y)\frac{2\bar{x}}{(x\bar{y}-\omega^2(q^2)+i\varepsilon)(1-\bar{x}y)}\nonumber\\
&&+r_\chi^{V_2}\Phi_{V1}(x)\Phi_{v2}(y)\frac{2y}{(x\bar{y}-\omega^2(q^2)+i\varepsilon)(1-\bar{x}y)}\Big\}\,,\\
A_3^{f,0}&=&\pi{\int}_{0}^{1}dxdy\alpha_s(q^2)\Big\{r_\chi^{V_1}\Phi_{v1}(x)\Phi_{V2}(y)\frac{2(1+\bar{y})}{\bar{y}(x\bar{y}-\omega^2(q^2)+i\varepsilon)}\nonumber\\
&&-r_\chi^{V_2}\Phi_{V1}(x)\Phi_{v2}(y)\frac{2(1+x)}{x(x\bar{y}-\omega^2(q^2)+i\varepsilon)}\Big\}\,,
\end{eqnarray}
and $A_{1,2}^{f,0}=0$. The non-vanishing transverse amplitudes are 
\begin{eqnarray}
A_1^{i,-}&=&\pi\frac{2m_1m_2}{m_B^2}{\int}_{0}^{1}dxdy\alpha_s(q^2)\Big\{\phi_{b1}(x)\phi_{b2}(y)\big[\frac{1+\bar{y}}{(1-\bar{x}y)(x\bar{y}-\omega^2(q^2)+i\varepsilon)}\nonumber\\
&&+\frac{\bar{x}\bar{y}^2}{(1-\bar{x}y)(x\bar{y}-\omega^2(q^2)+i\varepsilon)^2}-\frac{\bar{x}\bar{y}^2}{(1-\bar{x}y)^2(x\bar{y}-\omega^2(q^2)+i\varepsilon)}\nonumber\\
&&+\frac{\bar{x}}{(x\bar{y}-\omega^2(q^2)+i\varepsilon)^2}\big]\Big\}\,,\\
A_2^{i,-}&=&\pi\frac{2m_1m_2}{m_B^2}{\int}_{0}^{1}dxdy\alpha_s(q^2)\Big\{\phi_{a1}(x)\phi_{a2}(y)\big[\frac{1+x}{(1-\bar{x}y)(x\bar{y}-\omega^2(q^2)+i\varepsilon)}\nonumber\\
&&+\frac{x^2y}{(1-\bar{x}y)(x\bar{y}-\omega^2(q^2)+i\varepsilon)^2}-\frac{x^2y}{(1-\bar{x}y)^2(x\bar{y}-\omega^2(q^2)+i\varepsilon)}\nonumber\\
&&+\frac{y}{(x\bar{y}-\omega^2(q^2)+i\varepsilon)^2}\big]\Big\}\,,\\
A_3^{i,-}&=&\pi{\int}_{0}^{1}dxdy\alpha_s(q^2)\Big\{\frac{2m_1}{m_2}r_\chi^{V_2}\phi_{a1}(x)\phi_{2}^{\bot}(y)\frac{1}{(x\bar{y}-\omega^2(q^2)+i\varepsilon)(1-\bar{x}y)}\nonumber\\
&&-\frac{2m_2}{m_1}r_\chi^{V_1}\phi_{1}^{\bot}(x)\phi_{b2}(y)\frac{1}{(x\bar{y}-\omega^2(q^2)+i\varepsilon)(1-\bar{x}y)}\Big\}\,,
\end{eqnarray}
\begin{eqnarray}
A_3^{f,-}&=&\pi{\int}_{0}^{1}dxdy\alpha_s(q^2)\Big\{\frac{2m_1}{m_2}r_\chi^{V_2}\phi_{a1}(x)\phi_{2}^{\bot}(y)\frac{1}{\bar{y}(x\bar{y}-\omega^2(q^2)+i\varepsilon)}\nonumber\\
&&+\frac{2m_2}{m_1}r_\chi^{V_1}\phi_{1}^{\bot}(x)\phi_{b2}(y)\frac{1}{x(x\bar{y}-\omega^2(q^2)+i\varepsilon)}\Big\}\,,\\
A_1^{i,+}&=&\pi\frac{2m_1m_2}{m_B^2}{\int}_{0}^{1}dxdy\alpha_s(q^2)\Big\{\phi_{a1}(x)\phi_{a2}(y)\big[\frac{\bar{y}}{(1-\bar{x}y)(x\bar{y}-\omega^2(q^2)+i\varepsilon)}\nonumber\\
&&+\frac{xy\bar{y}}{(1-\bar{x}y)(x\bar{y}-\omega^2(q^2)+i\varepsilon)^2}-\frac{xy\bar{y}}{(1-\bar{x}y)^2(x\bar{y}-\omega^2(q^2)+i\varepsilon)}\nonumber\\
&&+\frac{1}{\bar{y}^2(x\bar{y}-\omega^2(q^2)+i\varepsilon)}+\frac{xy}{\bar{y}(x\bar{y}-\omega^2(q^2)+i\varepsilon)^2}\big]\Big\}\,,\\
A_2^{i,+}&=&\pi\frac{2m_1m_2}{m_B^2}{\int}_{0}^{1}dxdy\alpha_s(q^2)\Big\{\phi_{b1}(x)\phi_{b2}(y)\big[\frac{x}{(1-\bar{x}y)(x\bar{y}-\omega^2(q^2)+i\varepsilon)}\nonumber\\
&&+\frac{x\bar{x}\bar{y}}{(1-\bar{x}y)(x\bar{y}-\omega^2(q^2)+i\varepsilon)^2}-\frac{x\bar{x}\bar{y}}{(1-\bar{x}y)^2(x\bar{y}-\omega^2(q^2)+i\varepsilon)}\nonumber\\
&&+\frac{1}{x^2(x\bar{y}-\omega^2(q^2)+i\varepsilon)}+\frac{\bar{x}\bar{y}}{x(x\bar{y}-\omega^2(q^2)+i\varepsilon)^2}\big]\Big\}\,,
\end{eqnarray}
where $q^2\simeq x\bar{y}M_B^2$ is the time-like gluon momentum square. As found in Ref.~\cite{Chang1}, the hard-spectator scattering  contributions are real, but the annihilation contributions are complex with a large imaginary part. Their contributions are dominated by the value of the effective dynamical gluon mass scale $m_g$.

\begin{figure}[t]
\begin{center}
\subfigure[]{\includegraphics [width=6cm]{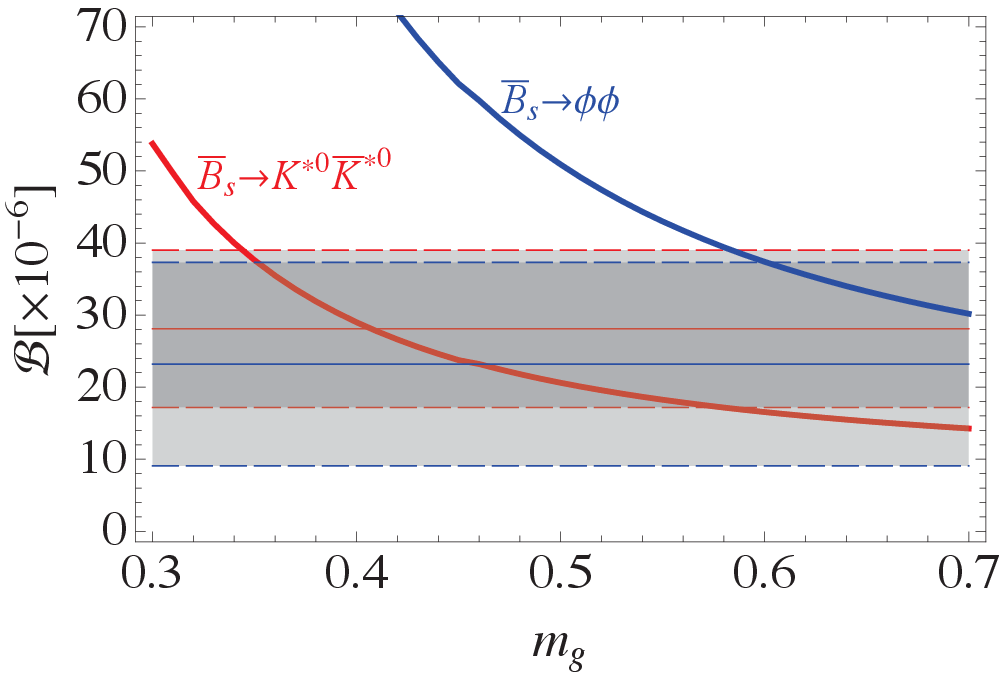}}\\
\subfigure[]{\includegraphics [width=6cm]{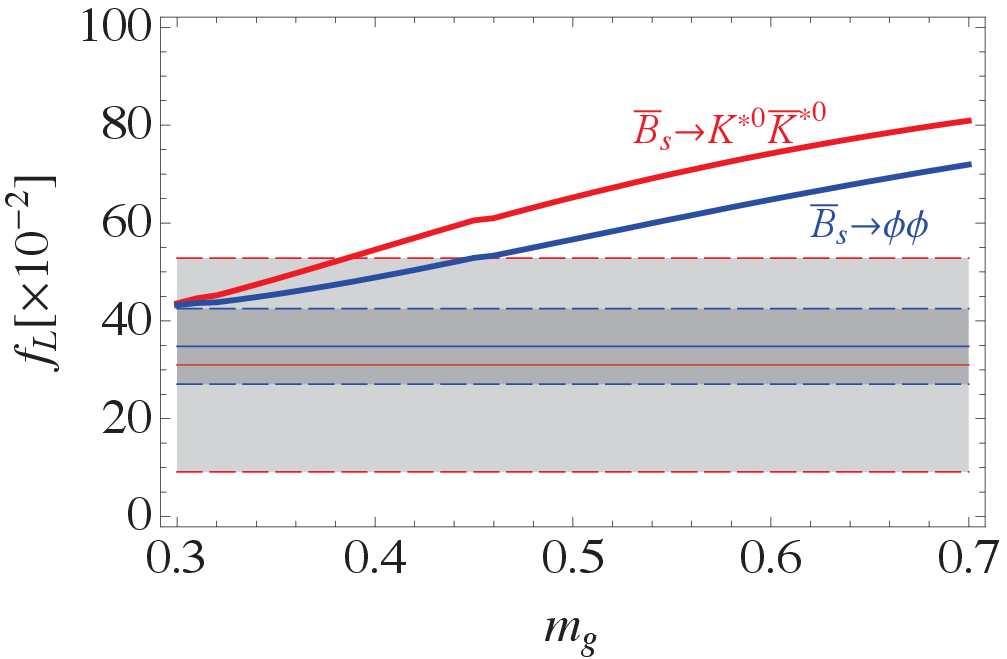}}~~~~~~~~~~
\subfigure[]{\includegraphics [width=6cm]{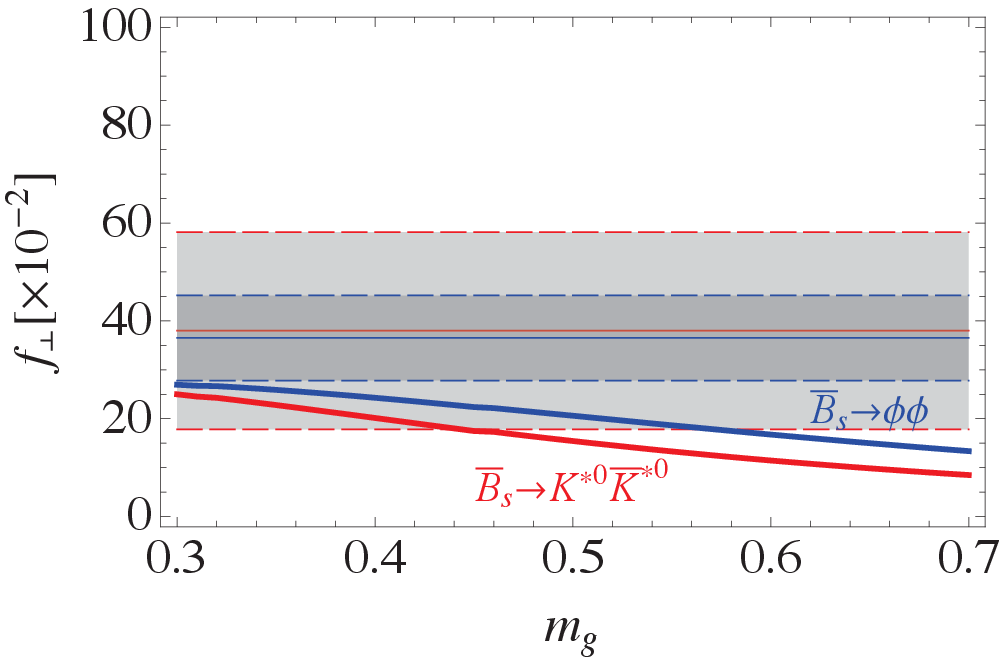}}
\centerline{\parbox{16cm}{\caption{\label{obsmg}\small The dependence of ${\cal B}(\bar{B}_s^0\to K^{\ast 0}\bar{K}^{\ast 0}\,,\phi\phi)$ and  $f_{L,\bot}(\bar{B}_s^0\to K^{\ast 0}\bar{K}^{\ast 0}\,,\phi\phi)$ on the effective dynamical gluon mass scale. The dashed lines correspond to the error bars~($1.68\sigma$).}}}
\end{center}
\end{figure}

With the default values of the input parameters, the dependence of ${\cal B}(\bar{B}_s^0\to K^{\ast 0}\bar{K}^{\ast 0}\,,\phi\phi)$ and  $f_{L,\bot}(\bar{B}_s^0\to K^{\ast 0}\bar{K}^{\ast 0}\,,\phi\phi)$ on the parameter $m_g$ is shown by Fig.~\ref{obsmg}. From Figs.~\ref{obsmg}~(b) and (c), we find that the QCDF predictions for $f_{L(\bot)}(\bar{B}_s^0\to K^{\ast 0}\bar{K}^{\ast 0}\,,\phi\phi)$ could be reduced (enhanced) to the experimental measurements with a small effective dynamical gluon mass $m_g\lesssim0.4{\rm GeV}$, which leads to a large annihilation contribution. Meanwhile, as Fig.~\ref{obsmg}~(a) shows, a small $m_g$ is also allowed by the constraint from ${\cal B}(\bar{B}_s^0\to K^{\ast 0}\bar{K}^{\ast 0})$. However, such small $m_g$ value would induce too large ${\cal B}(\bar{B}_s^0\to\phi\phi)$, which is much larger than the experimental data. 

With a default $m_g$ value $0.5{\rm GeV}$, we present our predictions for the observables in the last column of Table~\ref{VV}. We find that our prediction ${\cal B}(\bar{B}_s^0\to K^{\ast 0}\bar{K}^{\ast 0})=(20.6^{+6.5}_{-5.2})\times10^{-6}$ agrees well with the experimental data $(28.1\pm6.5)\times10^{-6}$. However, unfortunately, ${\cal B}(\bar{B}_s^0\to \phi\phi)$ is enhanced much by the annihilation corrections,  which lower limit $(49.9-13.3)\times10^{-6}$ is about $1.6\sigma$ larger than the CDF measurement $(23.2\pm8.4)\times10^{-6}$. In fact, with any $m_g$ value, as analysis in the last section, the theoretical prediction of ${\cal B}(\bar{B}_s^0\to \phi\phi)$ is always much larger than ${\cal B}(\bar{B}_s^0\to K^{\ast 0}\bar{K}^{\ast 0})$, which is also can be seen from Fig.~\ref{obsmg}~(a). Because a small $m_g$, which is help to accommodate the discrepency of $f_{L,\bot}(\bar{B}_s^0\to K^{\ast 0}\bar{K}^{\ast 0}\,,\phi\phi)$ between the theoretical predictions and experimental data as  Figs.~\ref{obsmg}~(b) and (c) showing, is excluded by the constraint from ${\cal B}(\bar{B}_s^0\to \phi\phi)$, the predictions of scheme~II for $f_{L(\bot)}(\bar{B}_s^0\to K^{\ast 0}\bar{K}^{\ast 0}\,,\phi\phi)$ are larger~(smaller) than the experimental measurements, which is similar to the results of scenarios $\overline{S4}$. For the $\bar{B}_s^0\to\rho\rho$ decays, their branching fractions in scheme~II are significantly larger than the ones in scheme~I. So, the upcoming measurements on ${\cal B}(\bar{B}_s^0\to \rho\rho)$ could give a judgment on the two schemes.

\section{Conclusion}
Motivated by the most recently observed large branching fraction of the pure annihilation decay $\bar{B}_s^0\to\pi^+\pi^-$ and large transverse polarization fractions in the $\bar{B}_s^0\to K^{\ast 0}\bar{K}^{\ast 0}\,,\phi\phi$ decays, we revisit the hard spectator and annihilation corrections in non-leptonic $\bar{B}_s^0$ decays within QCD factorization approach. In this paper, two schemes~(parameterization and using an infrared finite gluon propagator) to model the effects of  the end-point singularity in hard spectator and annihilation corrections are evaluated. In our numerical evaluations, all of the pure annihilation and well measured $\bar{B}_s^0$ decays are studied in detail simultaneously. Our main conclusions are summarized as:
\begin{itemize}
\item  For $\bar{B}_s^0\to PP$ decays, within scheme~I, due to the large ${\cal B}(\bar{B}_s^0\to\pi^+\pi^-)$ measured by CDF and LHCb collaborations, a large annihilation parameter $\rho_s^{PP}$ is required. Under the constraints from  ${\cal B}(\bar{B}_s^0\to\pi^+\pi^-\,,\pi^-K^+\,,K^-K^+)$  and  $A_{CP}(\bar{B}_s^0\to\pi^-K^+)$, two solutions $\rho_s^{PP}\sim2.5(3.5)$ and $\phi_s^{PP}\sim-84^{\circ}(116^{\circ})$ are found, which are significantly larger than these values $\rho^{PP}\sim1$ and $\phi^{PP} \sim-55^{\circ}$ adapted in the literature. With the obtained two solutions of the annihilation parameters given in Eq.~(\ref{SPPAB}), all of the the QCDF predictions for $\bar{B}_s^0\to PP$ decays agree well with the available experimental data. 
\item For $\bar{B}_s^0\to VV$ decays, within scheme~I, the measured small $f_{L}(\bar{B}_s^0\to K^{\ast 0}\bar{K}^{\ast 0}\,,\phi\phi)$ and large $f_{\bot}(\bar{B}_s^0\to K^{\ast 0}\bar{K}^{\ast 0}\,,\phi\phi)$ could be accommodated by the annihilation contributions. With the constraints from ${\cal B}(\bar{B}_s^0\to K^{\ast 0}\bar{K}^{\ast 0}\,,\phi\phi)$ and $f_{L\,,\bot}(\bar{B}_s^0\to K^{\ast 0}\bar{K}^{\ast 0}\,,\phi\phi)$, we find four solutions of the annihilation parameters $\rho_s^{VV}$ and $\phi_s^{VV}$ given by Eq.~(\ref{SVV}). Some of these solutions will be confirmed or refuted  by the upcoming LHCb measurement on $\bar{B}_s^0\to\rho\rho$ decays.
\item Within scheme~II, using the effective gluon mass $m_g=0.48{\rm GeV}$, QCDF predictions for $\bar{B}_s^0\to PP$ decays are found to be in good agreement with the available experimental results. Furthermore, some of its predictions are different from the ones in scheme~I, such as the branching fractions and direct CP asymmetries of $\bar{B}_s^0\to\rho\pi$ decays, which will be judged by the upcoming LHCb and proposed super-B experiments. For the $\bar{B}_s^0\to VV$ decays, $f_{L\,,\bot}(\bar{B}_s^0\to K^{\ast 0}\bar{K}^{\ast 0}\,,\phi\phi)$ could be accommodated by the annihilation contributions with a small $m_g\lesssim0.4{\rm GeV}$, which unfortunately is excluded by  ${\cal B}(\bar{B}_s^0\to\phi\phi)$. 
\item Within both scheme~I and II, ${\cal B}(\bar{B}_s^0\to\phi\phi)$ is always larger than ${\cal B}(\bar{B}_s^0\to K^{\ast 0}\bar{K}^{\ast 0})$, which significantly conflicts with the LHCb and CDF observation ${\cal B}(\bar{B}_s^0\to\phi\phi)\simeq {\cal B}(\bar{B}_s^0\to K^{\ast 0}\bar{K}^{\ast 0})$. A similar situation is also presented in pQCD approach as summarized  in Ref.\cite{Cheng2} and Eqs.~(\ref{QCDF1}, \ref{QCDF2}, \ref{pQCD}). Thus,  the present experimental result ${\cal B}(\bar{B}_s^0\to\phi\phi)\simeq {\cal B}(\bar{B}_s^0\to K^{\ast 0}\bar{K}^{\ast 0})$  rises a challenge to theoretical approaches for 
B nonleptonic decay. The further refined measurements and theoretical studies are required to resolve  such a possible anomaly.
\end{itemize}

\section*{Acknowledgments}
We are grateful to Hai-Yang Cheng and De-Shan Yang for helpful discussions. The work is supported by the National Natural Science Foundation under Grant Nos. 11075059 and 11105043, Research Fund for the Doctoral Program of Higher Education of China under Grant No. 20114104120002, and China Postdoctoral Science Foundation under Grant No. 2011M500118.

\begin{appendix}
\section*{AppendixA: Decay amplitudes with QCDF}
For selfconsitence of this paper, the following decay amplitudes are recapitulated from Refs.~\cite{Beneke2,Beneke3}. The amplitudes of $\bar{B}_s\to \pi\rho$ decays are
\begin{eqnarray}
{\cal A}_{\bar{B}_s\to \pi^+ \rho^-}&=& B_{\pi \rho} \Big[b_4^p - \frac{1}{2}b_{4,{\rm EW}}^p\Big]+B_{\rho \pi} \Big[\delta_{pu} b_1 + b_4^p + b_{4,{\rm EW}}^p\Big], \\
2{\cal A}_{\bar{B}_s\to \pi^0 \rho^0}&=& B_{\pi \rho} \Big[\delta_{pu} b_1 + 2b_4^p + \frac{1}{2}b_{4,{\rm EW}}^p\Big]+ B_{\rho \pi} \Big[\delta_{pu} b_1 + 2b_4^p + \frac{1}{2}b_{4,{\rm EW}}^p\Big].
\end{eqnarray}
The amplitude of $\bar{B}_s\to \pi^- \rho^+$ decay is obtained from the first expression by interchanging $(\pi) \leftrightarrow (\rho)$ everywhere. The expressions for the $\bar{B}_s\to \pi \pi$ and $\rho\rho$ amplitudes are obtained by setting $(\rho)\to (\pi)$ and $(\pi)\to (\rho)$, respectively.

The decay amplitudes of $\bar{B}_s\to \pi^- K^+\,,K^- K^+\,,$ and $\bar{K}^{\ast0} K^{\ast0}\,,\phi\phi$ decays are :
\begin{eqnarray}
{\cal A}_{\bar{B}_s\to \pi^- K^+}
   &=& A_{K \pi}\Big[\delta_{pu} \alpha_1 + \alpha_4^p + \alpha_{4,{\rm
    EW}}^p + \beta_3^p - \frac{1}{2}\beta_{3,{\rm EW}}^p\Big],\\
{\cal A}_{\bar{B}_s\to K^- K^+}
   &=& B_{K^- K^+} \Big[
    \delta_{pu} b_1 + b_4^p + b_{4,{\rm EW}}^p\Big]\nonumber\\
   &&+ A_{K^+ K^-}\Big[\delta_{pu} \alpha_1 + \alpha_4^p + \alpha_{4,{\rm
    EW}}^p + \beta_3 ^p+ \beta_4^p + \frac{1}{2}\beta_{3,{\rm EW}}^p - \frac{1}{2}\beta_{4,{\rm EW}}^p\Big],\\
\label{amp5_SM}
{\cal A}_{\bar{B}_s\to\bar{K}^{\ast0} K^{\ast0}}
   &=& B_{\bar{K}^{\ast0} K^{\ast0}} \Big[b_4^p - \frac{1}{2}b_{4,{\rm EW}}^p\Big]\nonumber\\
   &&+ A_{K^{\ast0} \bar{K}^{\ast0}}\Big[\alpha_4^p - \frac{1}{2}\alpha_{4,{\rm
    EW}}^p + \beta_3^p + \beta_4^p - \frac{1}{2}\beta_{3,{\rm EW}}^p - \frac{1}{2}\beta_{4,{\rm EW}}^p\Big],\\
\label{amp6_SM}
{\cal A}_{\bar{B}_s\to \phi \phi}
   &=& 2A_{\phi \phi}\Big[\alpha_3^p + \alpha_4^p - \frac{1}{2}\alpha_{3,{\rm
    EW}}^p - \frac{1}{2}\alpha_{4,{\rm EW}}^p + \beta_3^p + \beta_4^p - \frac{1}{2}\beta_{3,{\rm EW}}^p - \frac{1}{2}\beta_{4,{\rm EW}}^p\Big]\,.
\end{eqnarray}
The explicit expressions for the coefficients $\alpha_i^p\equiv\alpha_i^p(M_1M_2)$ and $\beta_i^p\equiv\beta_i^p(M_1M_2)$ can be found in Ref.~\cite{Beneke2,Beneke3}.

\section*{AppendixB: Theoretical input parameters}
For the CKM matrix elements, we adopt the CKMfitter Group's
fitting results~\cite{CKMfitter}
\begin{eqnarray}
\overline{\rho}=0.144\pm0.025\,,
\overline{\eta}=0.342^{+0.016}_{-0.015}\,,
A=0.812^{+0.013}_{-0.027}\,,
\lambda=0.22543\pm0.00077\,.
\end{eqnarray}

As for the quark mass, we take~\cite{PDG10,PMass,HPQCD:2006}
\begin{eqnarray}
 &&m_u=m_d=m_s=0, \quad m_c=1.61^{+0.08}_{-0.12}\,{\rm GeV},\nonumber\\
 &&m_b=4.78^{+0.21}_{-0.07}\,{\rm GeV}, \quad m_t=172.4\pm1.22\,{\rm GeV}.
\end{eqnarray}
for the pole masses and
\begin{eqnarray}
\frac{\overline{m}_s(\mu)}{\overline{m}_q(\mu)}&=&27.4\pm0.4\,,\quad
\overline{m}_{s}(2\,{\rm GeV})=87\pm6\,{\rm MeV},
\quad\overline{m}_{c}(\overline{m}_{c})=1.27^{+0.07}_{-0.09}\,{\rm GeV}\,\nonumber\\
\overline{m}_{b}(\overline{m}_{b})&=&4.19^{+0.18}_{-0.06}\,{\rm GeV}\,,\quad
\overline{m}_{t}(\overline{m}_{t})=164.8\pm1.2\,{\rm GeV}\,,
\end{eqnarray}
for the running masses, where $m_q=m_{u,d,s}$.

The decay constants are~\cite{PDG10,DecayCon,BallZwicky}
\begin{eqnarray}
f_{B_{s}}&=&(231\pm15)~{\rm MeV}\,,f_{B_{d}}=(190\pm13)~{\rm MeV}\,,\nonumber\\
f_{\pi}&=&(130.4\pm0.2)~{\rm MeV}\,,f_{K}=(156.1\pm0.8)~{\rm MeV}\,,f_{\rho}=(209\pm2)~{\rm MeV}\,.
\end{eqnarray}
As for the B-meson lifetimes,
$\tau_{B_{d}}=1.525\,{\rm ps}~\cite{PDG10}$ and $\tau_{B_{s}}=1.472\,{\rm ps}~\cite{PDG10}$
are used. Furthermore, the form factor $F^{B\to \pi}_{0}(0)=0.258\pm0.031$ ~\cite{BallZwicky} is used to evaluate the amplitude of $\bar{B}_d^0\to\pi^+K^-$ decay. For $\bar{B}_s^0\to \pi^-K^+$, $K^-K^+$ decays, we shall use $F^{B_s\to K}_{0}(0)=0.24$ obtained by both lattice and pQCD calculations and suggested by Ref.~\cite{Cheng2}. For the other decay constants and form factors related to $\bar{B}_s^0\to K^{\ast 0}\bar{K}^{\ast 0}$ and $\bar{B}_s^0\to\phi\phi$, we choose the similar inputs used in  Ref.~\cite{Beneke3}. In which, these values follow the QCD sum rule calculation~Ref.~\cite{BallZwicky}, but some modifications within theoretical errors are made to improve the description of data.  Their values are  
\begin{eqnarray}
&&f_{K^{\ast}}=(218\pm4)~{\rm MeV}\,,f_{K^{\ast}}^{\bot}(2{\rm GeV})=(175\pm25)~{\rm MeV}\,,\nonumber\\
&&f_{\phi}=(221\pm3)~{\rm MeV}\,,f_{\phi}^{\bot}(2{\rm GeV})=(175\pm25)~{\rm MeV}\,,\nonumber\\
&&A_0^{B_s\to \bar{K}^{\ast}}=0.38\pm0.03\,,F_-^{B_s\to \bar{K}^{\ast}}=0.53\pm0.05\,,F_+^{B_s\to \bar{K}^{\ast}}=0.00\pm0.06\nonumber\\
&&A_0^{B_s\to \phi}=0.38^{+0.10}_{-0.02}\,,F_-^{B_s\to \phi}=0.65^{+0.14}_{-0.00}\,,F_+^{B_s\to \phi}=0.00\pm0.06\,.
\end{eqnarray}

\end{appendix}

 \end{document}